\documentclass[
 reprint,
superscriptaddress,
 amsmath,amssymb,
 aps,
pra,
longbibliography
]{revtex4-2}

\usepackage{graphicx}
\usepackage{dcolumn}
\usepackage{bm}
\usepackage{bbm}
\usepackage[utf8]{inputenc}
\usepackage{amsfonts}
\usepackage{amsthm}
\usepackage{tikz}
\usepackage{physics}
\usepackage{comment}
\usepackage{xcolor}
\usepackage[caption=false,justification=justified]{subfig}
\usepackage{blindtext}
\usepackage{pgfplots,pgfplotstable}
\usepgfplotslibrary{polar}
\usepgfplotslibrary{colormaps}
\pgfplotsset{compat=newest}
\usepackage{hyperref}
\usepackage{xr-hyper}
\usepackage{footmisc}
\usepackage{cancel}
\usepackage{cleveref}
\crefname{equation}{Eq.}{Eqs.}

\renewcommand{\O}{\mathcal{O}}

\providecommand{\abs}[1]{\left\lvert#1\right\rvert}

\newcommand{\e}{\mathrm{e}}
\newcommand{\ii}{\mathrm{i}}

\newcommand{\Var}[1]{\mathrm{Var}[#1]}

\begin{document}
\title{Momentum-entangled two-photon interference for quantum-limited transverse-displacement estimation}

\author{Danilo Triggiani} \email{danilo.triggiani@port.ac.uk} \affiliation{School of Mathematics and Physics, University of Portsmouth, Portsmouth PO1 3HF, UK} \affiliation{Dipartimento di Fisica, Politecnico di Bari, Bari 70126, Italy}
\affiliation{INFN, Sezione di Bari, Bari 70126, Italy}
\author{Vincenzo Tamma} \email{vincenzo.tamma@port.ac.uk} \affiliation{School of Mathematics and Physics, University of Portsmouth, Portsmouth PO1 3HF, UK} \affiliation{Institute of Cosmology and Gravitation, University of Portsmouth, Portsmouth PO1 3FX, UK}

\date{\today}
\begin{abstract} We propose a scheme achieving the ultimate quantum precision for the estimation of the transverse displacement between two interfering photons. Such a transverse displacement could be caused, for example, by the refracting properties of the propagation medium or by the orientation of a system of mirrors. By performing transverse-momentum sampling interference between polarization-entangled pairs of photons that propagate with different momenta, we show that it is possible to perform transverse-displacement estimation with a precision that increases with the difference of the transverse momenta of the photons. We show that the precision achieved with our scheme is independent of the value of the displacement, useful when tracking a variable displacement. Moreover, only for small displacements, we show that the estimation can be performed without the need for transverse-momentum-resolving detectors. More fundamentally, we demonstrate that it is the quantum interference arising from two-photon entanglement in the transverse momenta at the very heart of the foreseen quantum-limited sensitivity in the spatial domain. \end{abstract}

\pacs{Valid PACS appear here}
\maketitle
\section{Introduction} Two-photon interference, e.g. observed when a pair of photons impinges on the two faces of a balanced beam splitter, is an established quantum phenomenon routinely employed for the development of novel quantum technologies~\cite{Hong1987, Shih1988, Bouchard2021}. When employing single-photon detectors at the outputs of the beam splitter, the rate at which both detectors click simultaneously depends on the differences between the quantum states of the two photons. While identical photons always `bunch' together, the rate of coincidences changes when gradually introducing differences between the two states of the photons, reproducing the well-known Hong-Ou-Mandel dip~\cite{Hong1987, Shih1988}, or quantum beats~\cite{Legero2003, Legero2004, Chen2023}. A typical application of this effect is thus found in sensing and metrology, where, for example, it has been employed for the estimation of optical lengths and time delays~\cite{Lyons2018, Scott2020, Fabre2021} or states of polarization~\cite{Harnchaiwat2020, Sgobba2023}. It has inspired sensing techniques such as quantum optical coherence tomography~\cite{Abouraddy2002, Nasr2009, Yepiz-Graciano2020, Hayama2022}, and applications have been proposed in fluorescence lifetime sensing~\cite{Lyons2023} and in single-molecule localization microscopy~\cite{Triggiani2024}. Furthermore, two-photon interference with non-degenerate SPDC has been employed for the estimation of small time delays~\cite{Chen2019}. On the other hand, employing detectors that resolve the inner modes of photons, such as frequency or transverse momentum, has been proven to drastically increase the dynamic range of sensing techniques based on two-photon interference while achieving optimal precision and simultaneously relieving the high-resolution requirements that are typical of high-precision direct measurements~\cite{Triggiani2023, Triggiani2024, Maggio2024, Muratore2024}. However, the uncertainty achievable in the estimation of transverse displacements through two-photon interference with separable photons is limited by the width of transverse-momentum distributions of the photons and may require large single-photon detector arrays~\cite{Triggiani2024}.
\begin{figure}[t] \centering \includegraphics[width=.95\columnwidth]{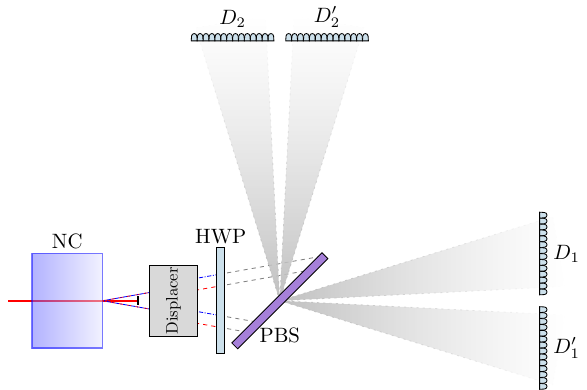} \caption{Schematic diagram of our transverse-momentum-resolving two-photon sensing scheme. Two SPDC polarization-entangled photons generated by pumping a type-II non-linear crystal (NC) propagate with different transverse momenta. The $H$ and $V$ polarization components of the two photons (drawn in red dashed and blue dot-dashed lines) are then transversally displaced by a quantity $\Delta x$. The two displaced photons are then let interfere, e.g. through a half-wave plate at 22.5$^\circ$ (HWP) and a polarizing beam splitter (PBS). Finally, single-photon cameras in the far field are used to simultaneously measure the transverse momenta $k'$ and $-k'$ of the photons and whether they end up at the same or opposite outputs of the PBS, sampling from the probability distribution in Eq.~\eqref{eq:Probs}. } \label{fig:Setup} \end{figure} 

Here we propose a transverse-momentum sampling two-photon interference technique to estimate the transverse displacement between two momentum-entangled photons, that we show can be generated through polarization-entangled type-II SPDC photons. The transverse displacement could be caused, e.g., by the refractive properties of a medium through which the two photons propagate or by a particular mirror configuration such as the one found in a tuneable beam displacer~\cite{Serrano2015}. Assessing the precision of our sensing scheme through Fisher information~\cite{Cramer1999, Rohatgi2000} and quantum Fisher information~\cite{Helstrom1969, Holevo2011} analysis, we show that the proposed scheme is optimal and does not require a displacement-dependent calibration, as it achieves the best precision possible in nature without prior knowledge and independently of the displacement to be estimated. Moreover, we show that such a precision increases with the difference in transverse momenta of the two entangled photons, instead of being limited by the much smaller width of the transverse-momentum distribution of each photon. The possibility to employ photons with narrow transverse-momentum distributions allows one to avoid using large single-photon detector arrays and instead to employ pairs of smaller cameras centered in the direction of propagation of each photon, as shown in \figurename~\ref{fig:Setup}. Finally, we also show that it is possible to replace the transverse-momentum-resolving detectors with bucket detectors for the measurement of small displacements, and we give estimates on the regimes of values of the displacement for which the non-resolving scheme retains its efficiency. Ultimately, we propose an intuitive description of the physics behind the quantum metrological advantage of the proposed technique.

\section{Preparation of the probe} 
We consider a polarization-entangled pair of quasi-monochromatic photons generated via type-II SPDC. Without loss of generality, we assume that the wave vectors of the pump and the downconverted photons lie on a single plane. In the remainder of this work, we will refer as transverse to any property or physical quantity related to the direction lying on this plane but transverse to the pump, and as longitudinal to properties related to the direction parallel to the pump. In the transverse direction, we write the two-photon state as 
\begin{equation} 
\ket{\psi_\mathrm{SPDC}} = \int\dd k_1\dd k_2\ g( k_1, k_2)\hat{a}_\mathrm{H}^\dag(k_1)\hat{a}_\mathrm{V}^\dag(k_2)\ket{0}, 
\label{eq:GenericPure} 
\end{equation} 
where $g(k_1,k_2)$ represents the joint amplitude distribution of the photonic transverse momenta, and $\hat{a}_{X}(k)$ denotes the bosonic operator associated with a photon with transverse momentum $k$ and polarization $X = H,V$. We now denote with $\Delta k/2$ and $-\Delta k/2$ the central transverse momenta of the two photons, and we assume that the difference $k_1-k_2$ of the transverse momenta of the two photons is narrowly distributed around the two values $k_1-k_2=\pm\Delta k$, namely, the two photons propagate non-collinearly. To satisfy this assumption, and recalling that the momentum conservation imposes $k_1+k_2=0$, we set 
\begin{multline} 
g(k_1, k_2)\equiv \mathcal{N}\delta( k_1+ k_2)\\ \times\left(f\left(\frac{ k_1- k_2-\Delta k}{2}\right)+f\left(\frac{k_2-k_1-\Delta k}{2}\right)\right) 
\label{eq:gDef} 
\end{multline} 
where $f(\cdot)$ is an amplitude probability distribution centered around zero and whose support is smaller than $\Delta k$ so that, $\forall k$, $f(k)f(-\Delta k-k)=0$, $\delta(\cdot)$ denotes the Dirac distribution, and $\mathcal{N}$ is a normalization constant.

Therefore, the state in Eq.~\eqref{eq:GenericPure} becomes
\begin{multline} 
\ket{\psi_\mathrm{SPDC}} = \mathcal{N}\int \dd k\ f(k)\Biggl[\hat{a}_\mathrm{H}^\dag\Biggl(\frac{\Delta k}{2}+k\Biggr)\hat{a}_\mathrm{V}^\dag\Biggl(-\frac{\Delta k}{2}-k\Biggr)\\ +\hat{a}_\mathrm{V}^\dag\Biggl(\frac{\Delta k}{2}+k\Biggr)\hat{a}_\mathrm{H}^\dag\Biggl(-\frac{\Delta k}{2}-k\Biggr)\Biggr]\ket{0}.
\label{eq:PolEnt} 
\end{multline} 
Notice that the particular choice of $g(k_1,k_2)$ in Eq.~\eqref{eq:gDef} gives rise to a polarization-entangled state in Eq.~\eqref{eq:PolEnt} of the type $\ket{HV}+\ket{VH}$, experimentally reproducible e.g. with a type-II non-collinear PPKTP crystal and a continuous-wave pump laser~\cite{Lee2016}. Since we are assuming that $f(k)f(-\Delta k-k)= 0$, the normalization of $\ket{\psi_{\mathrm{SPDC}}}$ implies ${\abs{\mathcal{N}}^2=1/2\delta(0)}$.

The two photons are then transversally displaced by an unknown and to be estimated quantity $\Delta x$. This can be introduced, for example, by means of a tunable beam displacer, as shown in \figurename~\ref{fig:Setup}, i.e. an optical device composed of a polarizing beam splitter and two mirrors~\cite{Serrano2015}, or a pair of opportunely aligned birefringent crystals, so that the horizontally polarized photon is displaced relatively to the vertically polarized one. Assuming that such a transverse displacement $\Delta x$ does not depend on the value of the transverse momentum of the photons, it causes a proportional relative phase in the transverse propagation of the two photons to the detectors, allowing us to write
\begin{multline} \!\!\!\!\!\ket{\psi_{\Delta x}}=\mathcal{N}\int \dd k\ f(k)\Biggl[\e^{\ii k \Delta x}\hat{a}_\mathrm{H}^\dag\Biggl(\frac{\Delta k}{2}+k\Biggr)\hat{a}_\mathrm{V}^\dag\Biggl(-\frac{\Delta k}{2}-k\Biggr)\\ +\e^{-\ii (k+\Delta k) \Delta x}\hat{a}_\mathrm{V}^\dag\Biggl(\frac{\Delta k}{2}+k\Biggr)\hat{a}_\mathrm{H}^\dag\Biggl(-\frac{\Delta k}{2}-k\Biggr)\Biggr]\ket{0},
\label{eq:PsiPar} 
\end{multline} 
where we omitted a global phase $\e^{\ii \Delta k\Delta x/2 }$. Mixing the modes $H$ and $V$ before the detection allows us to apply the transformation
\begin{equation} 
\hat{a}_H^\dag(\cdot)=\frac{1}{\sqrt{2}}(\hat{a}_1^\dag(\cdot)-\hat{a}^\dag_2(\cdot)),\ \hat{a}_V^\dag(\cdot)=\frac{1}{\sqrt{2}}(\hat{a}_1^\dag(\cdot)+\hat{a}^\dag_2(\cdot)),
\label{eq:Mixing} 
\end{equation} 
where $1,2$ label two output channels ultimately observed by the detectors. This can be done either by encoding the polarization state into separate spatial modes through a polarizing beam splitter and then letting them interfere at a balanced beam splitter or employing a half-wave plate at 22.5$^\circ$ and a polarizing beam splitter to mix the polarization modes into diagonal and anti-diagonal, as shown in \figurename~\ref{fig:Setup}. After the transformation in Eq.~\eqref{eq:Mixing}, the state in Eq.~\eqref{eq:PsiPar} can thus be written as 
\begin{equation} \ket{\psi_{\Delta x}}=\ket{\psi_A}+\ket{\psi_B}
\label{eq:State} 
\end{equation} 
where 
\begin{widetext} 
\begin{align} 
\label{eq:ComponentsState}
\begin{split} 
\ket{\psi_A}&=\frac{\mathcal{N}}{2}\int\dd k\ f( k)\left(\e^{\ii k\Delta x}-\e^{-\ii( k+\Delta k)\Delta x}\right)\Biggl[\hat{a}_1^\dag\Biggl( \frac{\Delta k}{2}+ k\Biggr)\hat{a}_2^\dag\Biggl( -\frac{\Delta k}{2}- k\Biggr)-\hat{a}_2^\dag\Biggl( \frac{\Delta k}{2}+ k\Biggr)\hat{a}_1^\dag\Biggl( -\frac{\Delta k}{2}- k\Biggr)\Biggr]\!\!\ket{0}\!,\\ \ket{\psi_B}&=\frac{\mathcal{N}}{2}\int\dd k\ f( k)\left(\e^{\ii k\Delta x}+\e^{-\ii( k+\Delta k)\Delta x}\right)\Biggl[\hat{a}_1^\dag\Biggl( \frac{\Delta k}{2}+ k\Biggr)\hat{a}_1^\dag\Biggl( -\frac{\Delta k}{2}- k\Biggr)-\hat{a}_2^\dag\Biggl( \frac{\Delta k}{2}+ k\Biggr)\hat{a}_2^\dag\Biggl( -\frac{\Delta k}{2}- k\Biggr)\Biggr]\!\!\ket{0} 
\end{split}
\end{align} 
\end{widetext} 
correspond to the event of the two photons ending up in different sides ($\ket{\psi_A}$) or in the same side ($\ket{\psi_B}$) of the beam splitter.
\section{Measurement scheme} 
\begin{figure}[t] 
\includegraphics[width=.95\columnwidth]{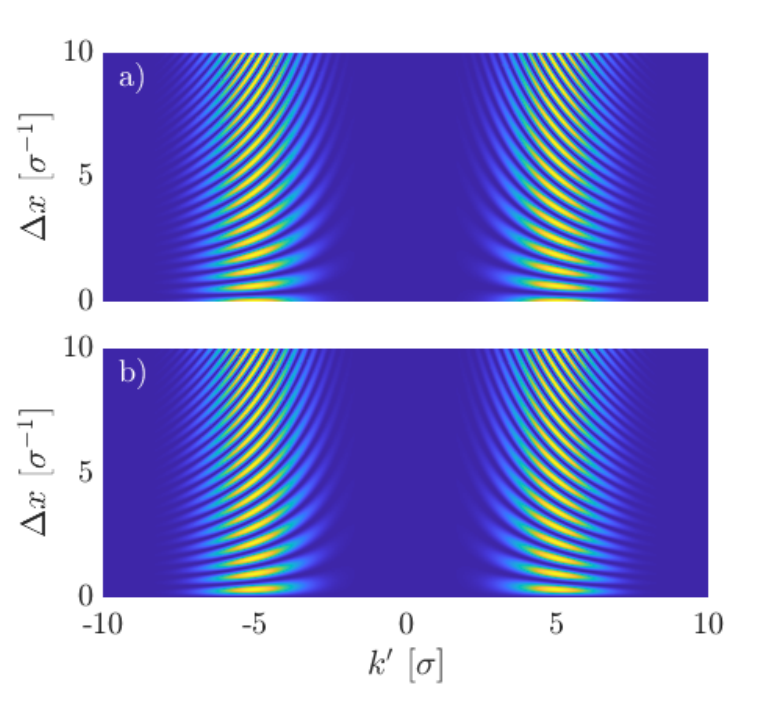} 
\caption{Plots of the probability $P_\gamma(k',X;\Delta x)$ in Eq.~\eqref{eq:Probs}, apart from the proportionality loss factor $(1-\gamma)^2$, to observe the transverse momenta $k'$ and $-k'$ of the two photons: a) in different ($X=A$) and b) in the same ($X=B$) output channel of the beam splitter, as a function of $k'$ and the unknown displacement $\Delta x$. Here, the transverse-momentum distribution $\abs{f}^2$ of the two photons is Gaussian with variance $\sigma^2=1$, which fixes a natural scale for $\Delta k=10\sigma$ and for $\Delta x$, while the loss factor $\gamma$ only appears in the proportionality factor $(1-\gamma)^2$. }
\label{fig:Probs} 
\end{figure}

Finally, the detectors measure the transverse momenta $k'$ and $-k'$ of the two photons, and they record whether the two photons end up in the opposite ($X=A$) or in the same ($X=B$) output channels of the beam splitter. From Eqs.~\eqref{eq:State}-\eqref{eq:ComponentsState} we evaluate in Appendix~\ref{app:Probs} the probability distributions $P_\gamma( k',X;\Delta x)$ associated with these outcomes, obtaining 
\begin{equation} 
\label{eq:Probs}
\begin{gathered} 
P_\gamma( k',X;\Delta x)=\frac{1}{4}(1-\gamma)^2C(k')\bigl(1+\alpha(X)\cos(2k'\Delta x)\bigr),\\ C(k')=\abs{f\left( k'- \frac{\Delta k}{2}\right)}^2+\abs{f\left( -\frac{\Delta k}{2}- k'\right)}^2 
\end{gathered} 
\end{equation} 
for $X\in\{A,B\}$, with $-\alpha(A)=\alpha(B)=1$, where $C(k')$ is an even, double-peaked envelope with peaks in $\pm\Delta k/2$, plotted in \figurename~\ref{fig:Probs} for a Gaussian amplitude distribution $\abs{f(x)}^2=\exp(-x^2/2\sigma^2)/\sqrt{2\pi\sigma^2}$ with variance $\sigma^2$, and we accounted for a loss probability $\gamma$ associated with the detection of each photon, which affects the probability simply through a proportionality factor. We have neglected the terms of the type $f( k'- \Delta k/2)f^*( -\Delta k/2- k')= 0$, i.e. the cross-terms between the two peaks centered in $\pm\Delta k/2$, as we are assuming that $\Delta k$ is larger than the width of the peaks. Noticeably, the probability $P_\gamma( k',X;\Delta x)$ oscillates with a period inversely proportional to $\Delta x$. The estimation of the displacement $\Delta x$ is then carried out by performing a number $N$ of sampling measurements of the values $(k', X)$, with a large enough resolution to resolve $\abs{f(k)}^2$ and the beating oscillations with period $\pi/\Delta x$ in Eq.\eqref{eq:Probs}. For example, assuming a Gaussian amplitude distribution $f$ with variance $\sigma^2$, and calling $\delta k$ the minimum variation of $k'$ measurable, the resolution conditions can be written as 
\begin{equation} 
\delta k \ll \sigma,\qquad \delta k\ll \frac{\pi}{\Delta x}.
\label{eq:Resolution} 
\end{equation} 
The set of outcomes $(k'_i, X_i)$, with $i =1,\dots, N$, is then employed to perform a standard maximum-likelihood estimation.

Since the probability distributions are concentrated around $k'=\pm \Delta k/2$, the range of transverse momenta of the two photons that need to be resolved by the detectors can be limited to the narrow peaks of $C(k')$ in Eq.~\eqref{eq:Probs}. Therefore, this scheme does not require detectors capable of resolving larger regions of transverse momenta, as it is instead customary in inner-mode-resolving two-photon interference sensing techniques employing separable photons, which need to have broad transverse-momentum distribution to increase the precision of the estimation~\cite{Triggiani2024}. For example, if the transverse momenta are measured through cameras detecting the photons with longitudinal component of the wave vector $\abs{K}$, propagating at a longitudinal distance $L$ in the far field, it is possible to employ one pair of cameras positioned at a transverse distance $d = \pm\frac{L\Delta k}{2\abs{K}}$ from the optical axes of each channel $1$ and $2$, instead of one large camera covering the whole transverse distance $2d$, as exemplified in \figurename~\ref{fig:Setup}.
\section{Quantum-limited Precision} We can assess the performance of our scheme for the estimation of the transverse distance $\Delta x$ in Eq.~\eqref{eq:PsiPar} by comparing the associated Cramér-Rao bound, i.e. the lowest uncertainty achievable with a given measurement scheme~\cite{Cramer1999, Rohatgi2000}, with the quantum Cramér-Rao bound, i.e. the ultimate uncertainty achievable regardless of the measurement scheme employed~\cite{Helstrom1969, Holevo2011}. These bounds respectively yield the inequalities 
\begin{equation} 
\delta_{\Delta x} \geqslant \frac{1}{\sqrt{N F_{\Delta x}}}\geqslant \frac{1}{\sqrt{N H_{\Delta x}}}, 
\label{eq:CRBs} 
\end{equation} 
where $\delta_{\Delta x}$ is the uncertainty associated with the estimation, $F_{\Delta x}$ is the Fisher information of our scheme, $H_{\Delta x}$ is the quantum Fisher information, and $N$ is the number of repetitions of the measurements. We show in Appendices~\ref{app:QFI} and~\ref{app:FI} that the Fisher information associated with our scheme is 
\begin{equation} 
F_{\Delta x} =(1-\gamma)^2 H_{\Delta x} =(1-\gamma)^2(\Delta k^2 + 4\sigma^2), 
\label{eq:FIs} 
\end{equation} 
i.e. our scheme is independent of the delay $\Delta x$ to be estimated, and it is optimal apart from a proportionality factor $(1-\gamma)^2$. The uncertainty achieved with our scheme is thus 
\begin{equation} 
\delta_{\Delta x} = \frac{1}{\sqrt{N(1-\gamma)^2 H_{\Delta x}}} = \frac{1}{\sqrt{N_\gamma(\Delta k^2 +4\sigma^2)}},
\label{eq:Precision} 
\end{equation} 
where $N_\gamma=N(1-\gamma)^2$ is the average number of measurements where both photons are detected. This means that the effect of losses can be simply countered by increasing the number of measurements. We notice that the Fisher information is independent of $\Delta x$, meaning that our scheme can be optimally employed to estimate in principle any separation induced between the two photons for any value of $\gamma$, provided that the transverse-momentum resolution $\delta k$ satisfies Eq.~\eqref{eq:Resolution}. Indeed, by resolving the transverse momenta of the photons, i.e. in the conjugate domain to the displacement $\Delta x$, the measurement `erases' the distinguishability between the photons at the detection, even for values of $\Delta x$ much larger than the width of their transverse spatial wavepackets.

We can see from Eq.~\eqref{eq:CRBs} that the precision achievable with our technique depends on the number $N_\gamma$ of observed pairs of photons and on the value of $\Delta k\gg\sigma$ that, in turn, can be maximized by optimizing the experimental conditions, such as the properties of the non-linear crystal, the frequency of the pump, and the geometry of the experimental setup. For example, one can engineer a value of the order of $\Delta k \simeq 2 k_z n \sin(\theta)\simeq 4.6*10^3\ \text{mm}^{-1}$~\cite{Coarse}, with $\Delta k\gg\sigma\simeq 2\pi*34\mathrm{mm}^{-1}\simeq0.2\mathrm{\mu m}^{-1}$ found e.g. in recent literature~\cite{Devaux2020}, with a wavelength of the emitted SPDC photons outside the crystal $\lambda = \frac{2\pi}{k_z} \simeq 800\text{nm}$, an index of refraction of the crystal $n\simeq 1.7$, and a cone aperture angle of the SPDC process equal to the angle of incidence on the surface of the crystal $\theta$ of the order of $\simeq 10^\text{o}$~\cite{LopezDuran2022}. In such an experimental scenario, the precision achievable by our scheme after $N_\gamma$ successful iterations is $\delta_{\Delta x}\simeq 1/\sqrt{N_\gamma\Delta k^2}\simeq (0.2/\sqrt{N_\gamma})\mathrm{\mu m}$ which, already for $N_\gamma\simeq 10^4$ pairs observed, is in the nanometer regime.

It is interesting to compare the result in Eq.~\eqref{eq:FIs} with the Fisher information $F_{\Delta x}^\mathrm{S}=2\sigma^2$ associated with the estimation of the displacement $\Delta x$ employing separable photons, where $\sigma^2$ here is the variance of the single-photon transverse-momentum distribution~\cite{Triggiani2024}. A simple physical picture that explains the difference between $F_{\Delta x}$ and $F_{\Delta x}^\mathrm{S}$ derives from the well-known result in quantum metrology stating that, for pure states where the quantity to be estimated can be thought of as generated through a unitary evolution, the quantum Fisher information is proportional to the variance of the generator $\hat{G}$ of the evolution~\cite{Giovannetti2006}. For the estimation of transverse displacements, such a generator is the semidifference of the transverse momenta of the two photons, i.e. $\hat{G}=(\hat{k}_1-\hat{k}_2)/2$, as we show in Appendix~\ref{app:Generator}. For two separable photons, whose transverse momenta are independent and identically distributed with variance $\sigma^2$, we easily evaluate $\Var{\hat{G}}=(\Var{\hat{k}_1}+\Var{\hat{k_2}})/4=2(\sigma^2/4)=\sigma^2/2$. For entangled photons, as described in Eq.~\eqref{eq:PolEnt}, whose transverse momenta are completely anti-correlated, $\hat{G}$ coincides with the single-photon transverse momentum since $\hat{k}_1=-\hat{k}_2$, so that $\Var{\hat{G}}= \Var{\hat{k}_1}=\sigma^2 + \Delta k^2/4$, where $\Delta k^2/4$ arises from the fact that each photon can equally likely propagate in opposite directions with an average transverse momentum $\pm\Delta k/2$.

\begin{figure} 
\centering
\includegraphics[width=1\columnwidth]{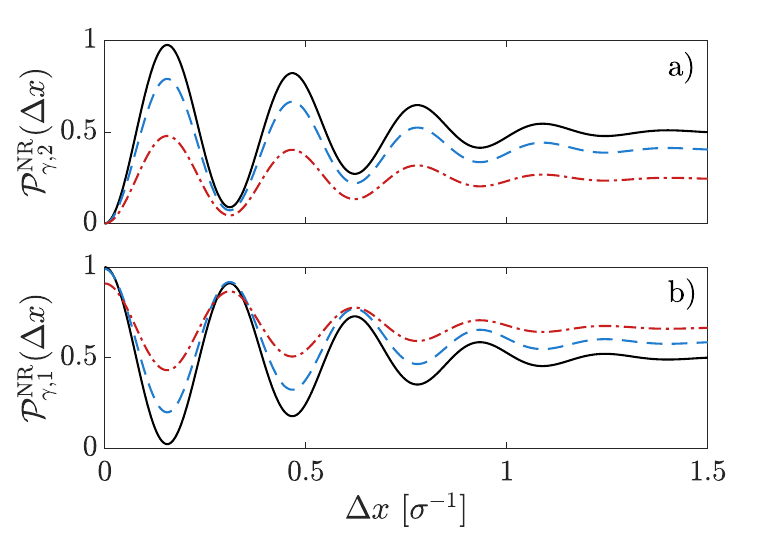} 
\caption{Plots of the probabilities $\mathcal{P}^{\mathrm{NR}}_{\gamma,i}$ in Eq.\eqref{eq:ProbNNR} for Gaussian transverse-momentum amplitude distribution $f(k')$ recording a) double click (${i=2}$) and b) single click ($i=1$) with on-off detectors as functions of $\Delta x$, for different values of the losses $\gamma$. For $\gamma = 0$ (black solid line), these correspond respectively to the probabilities $P^{\mathrm{NR}}_{\gamma}(A;\Delta x)$ and $P^{\mathrm{NR}}_{\gamma}(B;\Delta x)$ in Eq.~\eqref{eq:ProbNR} for number-resolving detectors, which are instead simply proportional to $(1-\gamma)^2$. In these plots, $\sigma$ fixes a natural scale for $\Delta k=20\sigma$ and for $\Delta x$ on the horizontal axes, while the loss factors are $\gamma=0$ (black solid lines), $\gamma=0.1$ (blue dashed lines), and $\gamma=0.3$ (red dot-dashed lines).} 
\label{fig:PNR} 
\end{figure}
\section{Non resolving the transverse momenta} 
We now show that it is possible to perform the estimation of small values of $\Delta x$ without resolving the transverse momenta of the photons. \subsection{Number-resolving detectors} If number-resolving detectors are employed, it is always possible to identify photon-loss events. In such a case, we have only access to the overall probabilities of bunching and coincidence 
\begin{equation} 
\label{eq:ProbNR} 
\begin{gathered} 
P^{\mathrm{NR}}_\gamma(X;\Delta x)=\frac{(1-\gamma)^2}{2}\left(1+\alpha(X)\chi(\Delta x)\right),\\ \chi(\Delta x) = \Re\left[\e^{\ii\Delta x\Delta k}\mathcal{F}_{\abs{f}^2}(2\Delta x)\right] 
\end{gathered} 
\end{equation} 
shown in \figurename~\ref{fig:PNR}, obtained in Appendix~\ref{app:NR} integrating Eq.~\eqref{eq:Probs} over all transverse momenta $k'$, where $\mathcal{F}_{\abs{f}^2}$ represents the Fourier transform of the frequency distribution probability $\abs{f}^2$, and $\Re[\cdot]$ denotes the real part. Notice that $P^{\mathrm{NR}}_\gamma(X;\Delta x)$ only depends on $\gamma$ through the proportionality factor $(1-\gamma)^2$.

We show in Appendix~\ref{app:NR} that the Fisher information $F^\mathrm{NR}_{\Delta x}$ associated with the non-resolved probability $P^\mathrm{NR}(X;\Delta x)$ in Eq.~\eqref{eq:ProbNR} can saturate the quantum Cramér-Rao bound in Eq.~\eqref{eq:CRBs} for lossless detectors $\gamma=0$ and $\Delta x\simeq 0$, a condition that in general can only be guaranteed after a prior calibration of the setup. For example, for a Gaussian transverse momentum distribution $\abs{f(k')}^2$, the Fisher information 
\begin{equation} 
F^\mathrm{NR}_{\Delta x} =(1-\gamma)^2\frac{(4\sigma^2\Delta x\cos(\Delta k\Delta x)+\Delta k\sin(\Delta k\Delta x))^2}{\e^{4\sigma^2\Delta x^2}-\cos^2(\Delta k\Delta x)}, 
\label{eq:FINR} 
\end{equation} 
evaluated in Appendix~\ref{app:NR} and shown in \figurename~\ref{fig:FINR} normalized over $F_{\Delta x}=(1-\gamma)^2H_{\Delta x}$, equates the quantum Fisher information in Eq.~\eqref{eq:FIs} for $\gamma=0$ and $\Delta x=0$ (neglecting the removable discontinuity). For $\Delta x\neq 0$ we show in Appendix~\ref{app:NR} that $F^\mathrm{NR}_{\Delta x}$ tends to the quantum Fisher information for $\tan(\Delta x \Delta k)\neq -4 \sigma^2 \Delta x/\Delta k$, a condition that excludes the values of $\Delta x$ for which $P^\mathrm{NR}(X;\Delta x)$ is locally independent of $\Delta x$, i.e. the stationary points, and $\sigma\Delta x \ll 1$, since 
\begin{equation} 
\frac{F^\mathrm{NR}_{\Delta x}}{F_{\Delta x}}\simeq 1 -4 \sigma^2\Delta x^2+\O(\Delta x^4). 
\label{eq:FINRsmall} 
\end{equation} 
Indeed, from \figurename~\ref{fig:FINR} and Eq.~\eqref{eq:FINR} we can see that the envelope of the Fisher information for the non-resolved approach decays as $\exp(-4\sigma^2\Delta x^2)$. This means that, assuming for example $\sigma\simeq 2\pi*34\mathrm{mm}^{-1}$ as in Ref.~\cite{Devaux2020}, for displacements such that $\Delta x \leqslant \frac{1}{6\sigma}\simeq 0.8\mathrm{\mu m}$, we have $F_{\Delta x}^\mathrm{NR}/F_{\Delta x}\simeq \mathrm{e}^{-1/9}\simeq 90\%$. However, for values such as $\Delta x \geqslant 2\mu$m, the Fisher information is substantially reduced by a factor~$\leqslant 1/\mathrm{e}\simeq 37\%$, pointing towards the need of resolving the transverse momenta in such a regime.

\begin{figure} 
\centering 
\includegraphics[width=\columnwidth]{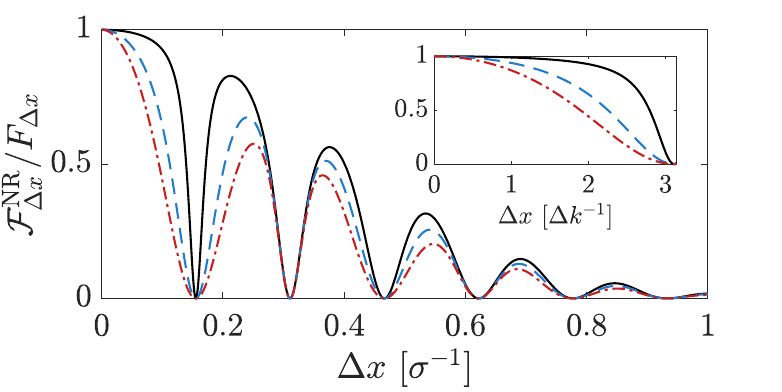} 
\caption{Plot of $\mathcal{F}^{\mathrm{NR}}_{\Delta x}$ in Eq.~\eqref{eq:FINNR} normalized over the Fisher information $F_{\Delta x}$ associated with the transverse-momentum resolving scheme for different values of the losses $\gamma$ and Gaussian photons. The plot for $\gamma=0$ (black solid line) also depicts the Fisher information $F^\mathrm{NR}_{\Delta x}$ in Eq.~\eqref{eq:FINR} of the number-resolving detection scheme since $\mathcal{F}^{\mathrm{NR}}_{\Delta x}=F^\mathrm{NR}_{\Delta x}$ for $\gamma=0$, and $F^\mathrm{NR}_{\Delta x}/F_{\Delta x}$ does not depend on $\gamma$. In the inset, we zoom in on the first peak of the Fisher information, rescaling the horizontal axis in units of $\Delta k^{-1}$. We can notice how $\mathcal{F}^{\mathrm{NR}}_{\Delta x}$ for $\gamma>0$ visibly decreases with $\Delta x$ in a few units of $1/\Delta k$, while $F^{\mathrm{NR}}_{\Delta x}$ remains mostly constant, as shown in \cref{eq:FINRsmall}. In these plots, $\sigma$ fixes a natural scale for $\Delta k=20\sigma$ and for $\Delta x$ on the horizontal axes, while the loss factors are $\gamma=0$ (black solid lines), $\gamma=0.1$ (blue dashed lines), and $\gamma=0.3$ (red dot-dashed lines).} 
\label{fig:FINR} 
\end{figure}
\subsection{On-off detectors} We now address a scenario where photon losses are not negligible. For measurements that resolve the transverse momenta, it is generally possible to identify the occurrence of losses with on-off detectors, given that events with two photons observed with the same transverse momenta are unlikely. On the other hand, when the transverse momenta are not resolved, losses can be identified only if the detectors employed are photon-number resolving. Instead, if the detectors do not resolve the number of photons, a coincidence event with a lost photon and a bunching event cause the same observable outcome, i.e. a single detector click. In this case, the outcome probabilities of observing $i$ clicks, with $i=1,2$, evaluated in Appendix~\ref{app:NR} and shown in \figurename~\ref{fig:PNR} are 
\begin{align} 
\label{eq:ProbNNR} 
\begin{split} 
\mathcal{P}^{\mathrm{NR}}_{\gamma,2}(\Delta x)&=\frac{(1-\gamma)^2}{2}\left(1-\chi(\Delta x)\right)=P_\gamma^\mathrm{NR}(A,\Delta x),\\ \mathcal{P}^{\mathrm{NR}}_{\gamma,1}(\Delta x)&=\frac{(1-\gamma)^2}{2}\left(\Gamma+\chi(\Delta x)\right)\geqslant P_\gamma^\mathrm{NR}(B,\Delta x) 
\end{split} 
\end{align} 
where the loss-dependent term $\Gamma = \frac{1+3\gamma}{1-\gamma}\geqslant 1$ implies that $\mathcal{P}^{\mathrm{NR}}_{\gamma,1}(\Delta x)\geqslant P_\gamma^\mathrm{NR}(B,\Delta x)$ as expected, given that single-click observations now include also possible additional bunching and coincidence events where one photon was lost. Noticeably, the behavior of the Fisher information $\mathcal{F}^{\mathrm{NR}}_{\Delta x}\leqslant F^{\mathrm{NR}}_{\Delta x}$ associated with the probabilities in Eq.~\eqref{eq:ProbNNR} for on-off detectors, when specialized to a Gaussian transverse momentum distribution $f(k')$, is similar to $F^\mathrm{NR}_{\Delta x}$ in Eq.~\eqref{eq:FINR}, and reads
\begin{equation} 
\mathcal{F}^{\mathrm{NR}}_{\Delta x} = \frac{(1-\gamma^2)(4\sigma^2\Delta x\cos(\Delta k\Delta x)+\Delta k\sin(\Delta k\Delta x))^2}{(\e^{2\sigma^2\Delta x^2}-\cos(\Delta k\Delta x))(\e^{2\sigma^2\Delta x^2}\Gamma+\cos(\Delta k\Delta x))}, 
\label{eq:FINNR} 
\end{equation} 
shown in \figurename~\ref{fig:FINR}, with $F^{\mathrm{NR}}_{\Delta x}=\mathcal{F}^{\mathrm{NR}}_{\Delta x}$ for $\gamma=0$. Interestingly, each individual fringe in the oscillations of $\mathcal{F}^{\mathrm{NR}}_{\Delta x}$ appears thinner than the ones of $F^\mathrm{NR}_{\Delta x}$ due to the presence of the factor $\Gamma$. This means that a more precise calibration is required to approximately reach each peak. In particular, we show in Appendix~\ref{app:NR} that, for $\gamma\neq 0$, it is possible to approximate the first peak of Fisher information associated with non-resolving on-off detectors to the one of the resolving scheme, i.e. $\mathcal{F}^{\mathrm{NR}}_{\Delta x}\simeq F_{\Delta x}$, only for $\Delta x\Delta k\ll 1$, since 
\begin{equation} 
\frac{\mathcal{F}^{\mathrm{NR}}_{\Delta x}}{F_{\Delta x}} \simeq 1 - \frac{\gamma}{2(1+\gamma)}\Delta k^2\Delta x^2+\O(\sigma^2\Delta x^2)+ \O(\Delta x^4). 
\label{eq:FINNRsmall} 
\end{equation} 
Therefore, when employing on-off detectors only, the values of the displacement $\Delta x\ll 1/\Delta k$ (of the order of $ 0.2\mathrm{\mu m}$ for $\Delta k\simeq 4.6*10^3\mathrm{mm}^{-1}$), can be estimated efficiently since they fall within the first peak of $\mathcal{F}^\mathrm{NR}_{\Delta x}$. The larger is the attainable value of $\Delta k$. the larger is the achievable precision in the estimation, but unfortunately the smaller is the range of values of $\Delta x$ that can be estimated efficiently with non-resolved measurements with on-off detectors
\subsection{Ambiguity of the estimation} The non-resolved probabilities in \cref{eq:ProbNR,eq:ProbNNR} present oscillations in the unknown displacement $\Delta x$ of period $\simeq 2\pi/\Delta k$ that make $P^\mathrm{NR}(X;\Delta x)$ and $\mathcal{P}^{\mathrm{NR}}(X;\Delta x)$ not injective as a function of $\Delta x$ (see \figurename~\ref{fig:PNR}), thus rendering the inversion problem to estimate $\Delta x$ through detectors that do not resolve the transverse momenta ambiguous irrespective of the attainable precision. This implies that, in order to retrieve the value of $\Delta x$ from the simple observation of the coincidence and bunching rates, one needs to know in advance the invertibility interval of the probabilities as functions of $\Delta x$ in which the true value of the delay lies, i.e. a prior knowledge of the value of $\Delta x$ of the order of $\pi/\Delta k$ is required, corresponding to $\simeq 0.7\mathrm{\mu m}$ if $\Delta k \simeq 4.6*10^3 \mathrm{mm}^{-1}$ as previously estimated. In other words, while the transverse-momentum-resolving approach is ambiguity-free and thus able to live-track large changes of the unknown displacement $\Delta x$ without the need of a prior calibration of the setup, the non-resolving measurement scheme requires some prior knowledge on $\Delta x$ and generally a calibration of the setup for an unambiguous estimation of the parameter.
\subsection{Discussing the optimality for small separations} Finally, we propose a heuristic approach to develop a better insight on the role of the transverse-momentum-resolving detection and to understand why a non-resolving technique is optimal only for small separations $\Delta x$, examining the structure of Eq.~\eqref{eq:ComponentsState}. Indeed, we can notice the presence of interfering terms that, at the detection of two photons with transverse momenta $k'$ and $-k'$, i.e. $\hat{a}^\dag_i(k')\hat{a}^\dag_j(-k')\ket{0}$ for $i,j\in\{1,2\}$, are of the type $\e^{-\ii\Delta k\Delta x/2}(\e^{\ii k'\Delta x}\pm \e^{-\ii k'\Delta x})$. These $k'$-dependent interference terms naturally arise from the coherent superposition, due to the mixing of the photons shown in Eq.~\eqref{eq:Mixing}, of two equally possible but indistinguishable events, where the photon displaced by a transverse length $\Delta x$ can either have transverse momentum $k'$ or $-k'$. By employing detectors that do not resolve the transverse momenta of the photons, these interference terms are averaged out over $k'$, together with the information they carry about $\Delta x$. However, for $\Delta x\simeq 0$, these $k'$-dependent interference terms only yield a negligible additional observable information of order $\mathcal{O}(\Delta x^2)$ on $\Delta x$, since $|\e^{\ii k'\Delta x}\pm \e^{-\ii k'\Delta x}|^2=(2\pm 2)+\mathcal{O}(\Delta x^2)$, with the dependence of $k'$ residing only in $\O(\Delta x^2)$.
\section{Conclusons} We have presented a sensing technique for the estimation at the ultimate quantum precision of the transverse displacement induced on two beams of momentum-entangled producible with polarization-entangled type-II SPDC. In particular, the sensitivity of our scheme increases with the difference in the transverse momenta of the two entangled photons. Compared to transverse-momentum-resolving techniques that employ separable photons, this offers a noticeable advantage since, in the latter, the precision of the estimation can only be increased by employing photons with larger variances in the transverse-momentum single-photon distributions. The sensitivity achievable with our scheme is also independent of the value of the displacement to be estimated, which renders the estimation calibration-free and effective when employed to live-track large variations of the unknown displacement.

We have also proposed a secondary approach that employs detectors that do not resolve the transverse momenta of the photons, analyzing the two distinct scenarios when number-resolving detectors or on-off detectors are operated. We have shown that both approaches achieve the ultimate precision, although only in the regime of small displacements, therefore requiring some prior knowledge on the unknown displacement and a prior calibration of the setup. In the momenta-resolved scenario, it is possible to identify the losses without the need of resolving the photon numbers, given that it is unlikely that both bunching photons are observed with the same transverse momenta. On the other hand, we have shown that number-resolved detectors are in general advantageous in the scenario where the transverse momenta are not resolved. Indeed, we have shown that if on-off detectors are used, the presence of losses has a higher detrimental effect on the precision. Furthermore, in such a case it is necessary to have an initial adaptation of the interferometer, which requires prior knowledge of the value of the displacement with precision of the order of the inverse of the transverse momenta difference of the entangled photons. On the other hand, by resolving the transverse momenta, no adaptation of the interferometer is required, and one can increase the precision by increasing the difference in the transverse momenta without affecting the range of values that it is possible to estimate.

This technique could possibly find practical applications in the analysis of the optical properties of birefringent materials or high-precision measurements of the orientation or the rotation of systems, since the relative displacement introduced by certain tunable beam displacers depends on the orientation of the device~\cite{Serrano2015}. On a more fundamental level, we have shown that the momentum-entangled photons generated with type-II SPDC can be employed to gain metrological advantage compared to similar schemes employing separable photons, and we have proposed a potential experimental scheme that, as a proof of principle, achieves such an advantage.

\section*{Acknowledgements}
VT acknowledges support from the Air Force Office of Scientific Research under award number FA8655-23-1-7046.
DT acknowledges the Italian Space Agency (ASI, Agenzia Spaziale Italiana) through the project ‘Subdiffraction Quantum Imaging’ (SQI) n. 2023-13-HH.0.
\appendix \onecolumngrid
\section{Evaluating the probabilities in \cref{eq:Probs}} \label{app:Probs}
Here, we evaluate the probability $P(k',X;\Delta x)$ in Eq.~\eqref{eq:Probs} to observe two photons with transverse momenta $k'$ and $-k'$ in either in opposite ($X=A$) or in the same ($X=B$) output channels of the beam splitter. We will first consider the lossless scenario and then use that result to implement a photon-loss probability $\gamma$.
\subsection{Lossless scenario, $\gamma = 0$} From Eqs.~\eqref{eq:State}-\eqref{eq:ComponentsState}, we can easily evaluate the probability to observe the two photons in opposite ports with transverse momenta $k'$ and $-k'$ as \begin{align} P(k',A;\Delta x) &= \abs{\bra{0}\hat{a}_1(k')\hat{a}_2(-k')\ket{\psi_{\Delta x}}}^2 = \abs{\bra{0}\hat{a}_1(k')\hat{a}_2(-k')\ket{\psi_A}}^2\notag \\ &=\frac{1}{8}\abs{\int\dd k\ f( k)\left(\e^{\ii k\Delta x}-\e^{-\ii( k+\Delta k)\Delta x}\right)\left[\delta\left(k'-\frac{\Delta k}{2}-k\right)-\delta\left(k'+\frac{\Delta k}{2}+k\right)\right]}^2\notag\\ &=\frac{1}{8}\abs{\e^{-\ii\frac{\Delta k}{2}}\left[f\left(k'-\frac{\Delta k}{2}\right)\left(\e^{\ii k'\Delta x}-\e^{-\ii k'\Delta x}\right) -f\left(-k'-\frac{\Delta k}{2}\right)\left(\e^{-\ii k'\Delta x}-\e^{\ii k'\Delta x}\right)\right]}^2\notag\\ &=\frac{1}{4}\left(\abs{f\left(k'-\frac{\Delta k}{2}\right)}^2+\abs{f\left(-k'-\frac{\Delta k}{2}\right)}^2\right)(1-\cos(2k'\Delta x)), \end{align} while the probability to observe the two photons in the same port with transverse momenta $k'$ and $-k'$ reads \begin{align} P(k',B;\Delta x) &= \frac{1}{2}\left(\abs{\bra{0}\hat{a}_1(k')\hat{a}_1(-k')\ket{\psi_B}}^2+\abs{\bra{0}\hat{a}_2(k')\hat{a}_2(-k')\ket{\psi_B}}^2\right)\notag\\ &=\frac{1}{8}\abs{\int\dd k\ f( k)\left(\e^{\ii k\Delta x}+\e^{-\ii( k+\Delta k)\Delta x}\right)\left[\delta\left(k'-\frac{\Delta k}{2}-k\right)+\delta\left(k'+\frac{\Delta k}{2}+k\right)\right]}^2\notag\\ &=\frac{1}{8}\abs{\e^{-\ii\frac{\Delta k}{2}}\left[f\left(k'-\frac{\Delta k}{2}\right)\left(\e^{\ii k'\Delta x}+\e^{-\ii k'\Delta x}\right) +f\left(-k'-\frac{\Delta k}{2}\right)\left(\e^{-\ii k'\Delta x}+\e^{\ii k'\Delta x}\right)\right]}^2\notag\\ &=\frac{1}{4}\left(\abs{f\left(k'-\frac{\Delta k}{2}\right)}^2+\abs{f\left(-k'-\frac{\Delta k}{2}\right)}^2\right)(1+\cos(2k'\Delta x)), \end{align} which coincides with the expression found in Eq.~\eqref{eq:Probs} in the main text for $\gamma=0$, once defined $C(k')$ as shown in the second line of Eq.~\eqref{eq:Probs}.
\subsection{Lossy scenario, $\gamma>0$}
To introduce the effect of losses, we simply observe that the only relevant events for the estimation of the separation $\Delta x$ are the events where both photons are observed. Indeed, we can model the presence of losses with beam splitters with reflectivity $\sqrt{\gamma}$ positioned, without loss of generality, before the detectors, transforming the terms in Eq.~\eqref{eq:ComponentsState} via \begin{equation} \hat{a}_i^\dag(k')\hat{a}_j^\dag(-k')\rightarrow (1-\gamma)\hat{a}_i^\dag(k')\hat{a}_j^\dag(-k')+\gamma \hat{e}_i^\dag(k')\hat{e}_j^\dag(-k')+\sqrt{\gamma(1-\gamma)}(\hat{a}_i^\dag(k')\hat{e}_j^\dag(-k')+\hat{e}_i^\dag(k')\hat{a}_j^\dag(-k')) ,\ i,j=1,2, \label{eq:TranformLoss} \end{equation} that is a sum of three terms corresponding to zero, one, and both photons lost in `environment' modes $\hat{e}_i(k')$. The probability to observe zero photons is thus given by the third term in Eq.~\eqref{eq:TranformLoss} \begin{equation} P_0= \sum_{i,j=1}^2\int\dd k'\ \abs{\bra{0}\hat{e}_i(k')\hat{e}_j(-k')\ket{\psi_{\Delta x}}}^2 = \gamma^2\sum_{X=A,B}\int\dd k'\ P(k',X;\Delta x) = \gamma^2, \label{eq:P0} \end{equation} while the probability to observe one photon with transverse momentum $k'$ in the output port $i$ is given by the second term in Eq.~\eqref{eq:TranformLoss} \begin{equation} P_i(k')=\sum_{j=1,2}\abs{\bra{0}\hat{a}_i(k')\hat{e}_j(-k')\ket{\psi_{\Delta x}}}^2 = \gamma(1-\gamma)\sum_{X=A,B}P(k',X;\Delta x) = \frac{1}{2}\gamma(1-\gamma) C(k'), \label{eq:P1} \end{equation} Noticeably, both $P_0$ and $P_i(k')$ are independent of $\Delta x$, hence not yielding any contribution to the sensitivity of the scheme and thus can be neglected. Finally, the probability to observe both photons including the losses is similarly obtained by the first term in Eq.~\eqref{eq:TranformLoss} \begin{equation} P_\gamma(k',X;\Delta x) = (1-\gamma)^2P(k',X;\Delta x), \label{eq:P2} \end{equation} as shown in Eq.~\eqref{eq:Probs} in the main text. Clearly, the probabilities $P_0$, $P_i(k')$ and $P_\gamma(k',X;\Delta x)$ are correctly normalized, i.e. \begin{equation} P_0+\sum_{i=1,2}\int\dd k'\ P_i(k')+ \sum_{X=A,B}\int\dd k'\ P_\gamma(k',X;\Delta x) =\gamma^2+2\gamma(1-\gamma)+(1-\gamma)^2=1. 
\end{equation}
\section{Evaluation of the quantum Fisher information $H_{\Delta x}$} \label{app:QFI}
We here evaluate the quantum Fisher information $H_{\Delta x}$. We first introduce the expression of the quantum Fisher information valid for pure states, \begin{equation} H_{\Delta x} = 4(\braket{\partial\psi}-\abs{\braket{\psi}{\partial\psi}}^2), \label{eq:QFIDef} \end{equation} where $\partial$ represents the differentiation with respect to the unknown parameter $\Delta x$ to be estimated~\cite{Holevo2011}. We first evaluate \begin{equation} \ket{\partial\psi}=\mathcal{N}\int \dd k\ f( k)\left(\ii k\e^{\ii\Delta x k}\hat{a}_1^\dag\left( \frac{\Delta k}{2}+ k\right)\hat{a}_2^\dag\left( -\frac{\Delta k}{2}- k\right)-\ii( k+\Delta k)\e^{-\ii \Delta x( k+\Delta k)}\hat{a}_1^\dag\left( -\frac{\Delta k}{2}- k\right)\hat{a}_2^\dag\left( \frac{\Delta k}{2}+ k\right)\right), \end{equation} then we derive \begin{multline} \braket{\partial\psi}=\frac{1}{2}\int\dd k\dd k'\ f( k)f^*( k')\left( k k'\e^{\ii\Delta x( k- k')}+( k+\Delta k)( k'+\Delta k)\e^{-\ii\Delta x( k- k')}\right)\delta( k- k')\\ =\frac{1}{2}\int\dd k\ \abs{f( k)}^2( k^2+( k+\Delta k)^2)=\sigma^2+\frac{\Delta k^2}{2}, \label{eq:QFIP1} \end{multline} where $\sigma^2$ is the variance of the transverse momentum distribution $\abs{f(k)}^2$ and \begin{equation} \braket{\psi}{\partial\psi}=\frac{1}{2}\int\dd k\dd k'\ f( k)f^*( k')\left(\ii k\e^{\ii\Delta x( k- k')}-\ii( k+\Delta k)\e^{-\ii\Delta x( k- k')}\right)\delta( k- k')=-\frac{\ii}{2}\Delta k, \label{eq:QFIP2} \end{equation} where we once again made use of the condition on $\Delta k$ much larger than the support of $f( k')$. Plugging Eqs.~\eqref{eq:QFIP1}-\eqref{eq:QFIP2} into Eq.~\eqref{eq:QFIDef}, we obtain \begin{equation} H_{\Delta x}=4\sigma^2+\Delta k^2. \label{eq:QFIResult} \end{equation} which is the quantum Fisher information shown in the main text in Eq.~\eqref{eq:FIs}.

\section{Evaluation of the Fisher information $F_{\Delta x}$ in Eq.~\eqref{eq:FIs} of the transverse momentum resolving scheme} \label{app:FI} We can now evaluate the Fisher information for the estimation of $\Delta x$ associated with our scheme employing the definition~\cite{Cramer1999} \begin{equation} F_{\Delta x} = \mathbb{E}\left[(\partial\log P_\gamma( k',X;\Delta x))^2\right], \end{equation} where $\mathbb{E}[\cdot]$ represents the expectation value associated with the same probability distribution $P_\gamma(k',X;\Delta x)$ given in Eq.~\eqref{eq:Probs} in the main text. After some simple algebra, we obtain \begin{align} F_{\Delta x}&=\int\dd k'\ P_\gamma( k',A;\Delta x)(\partial\log P_\gamma( k',A;\Delta x))^2 + \int\dd k\ P_\gamma( k',B;\Delta x)(\partial\log P_\gamma( k',B;\Delta x))^2\notag\\ &=\frac{(1-\gamma)^2}{2}\int\dd k'\ \left(\abs{f\left( k'- \frac{\Delta k}{2}\right)}^2+\abs{f\left( -\frac{\Delta k}{2}- k'\right)}^2\right) (2 k')^2=4(1-\gamma)^2\left(\sigma^2+ \frac{\Delta k^2}{4} \right)\notag\\ &=(1-\gamma)^2 (4\sigma^2+\Delta k^2)\equiv (1-\gamma)^2 H_{\Delta x}. \end{align} which coincides with the expression of $F_{\Delta x}$ given in Eq.~\eqref{eq:FIs} in the main text.
\section{Quantum Fisher information as variance of the generator $\hat{G}=(\hat{k}_1-\hat{k}_2)/2$} \label{app:Generator}
We can easily see that any two-photon state of the type \begin{equation} \ket{\psi_{\Delta x}} = \int \dd k_1\dd k_2\ g(k_1,k_2)\e^{\ii (k_1 x_1+ k_2 x_2)} \hat{a}^\dag_\mathrm{H}(k_1)\hat{a}^\dag_\mathrm{V}(k_2)\ket{0} \equiv \e^{\ii \Delta x\, \hat{G}}\ket{\psi_{\Delta x=0}}, \label{eq:GeneratorPsi} \end{equation} with $x_1=(X_c+\Delta x)/2$ and $x_2=(X_c-\Delta x)/2$, can be rewritten through the generator $\hat{G}=(\hat{k}_1-\hat{k}_2)/2$ given by the semi-difference of the transverse momenta of the two photons with, for example, $\hat{k}_1\hat{a}_\mathrm{H}(k)=k \hat{a}_\mathrm{H}(k)$ and $\hat{k}_2\hat{a}_\mathrm{V}(k)=k \hat{a}_\mathrm{V}(k)$. Employing famous results of quantum metrology, we can evaluate the quantum Fisher information of the state in Eq.~\eqref{eq:GeneratorPsi} through the variance of the generator $\hat{G}$ on the state $\ket{\psi_{\Delta x=0}}$~\cite{Giovannetti2006}. For a separable state of two identical photons with a variance of the transverse momentum distribution $\sigma^2$, $g(k_1,k_2)$ is factorizable, $\hat{k}_1$ and $\hat{k}_2$ are independent, hence $H_{\Delta x} = 4\frac{\Var{\hat{k}_1}+\Var{\hat{k}_2}}{4}=2\sigma^2$. For entangled photons, $g(k_1,k_2)\equiv h(k_1)\delta(k_1+k_2)$, $\hat{k}_1=-\hat{k}_2$, hence $\hat{G}=\hat{k}_1$ and $H_{\Delta x}=4\Var{\hat{k}_1}$. If $\abs{h(k)}^2=(\abs{f(k-\Delta k/2)}^2+\abs{f(-k-\Delta k/2)}^2)/2$ has two peaks centered in $\pm\Delta k/2$ as in $\ket{\psi_{\Delta x=0}}=\ket{\psi_{\mathrm{SPDC}}}$ Eq.~\eqref{eq:gDef}, we have \begin{equation} H_{\Delta x}=4\Var{\hat{k}_1}=\frac{4}{2}\int\dd k \abs{f(k)}^2\left(\left(k-\frac{\Delta k}{2}\right)^2+\left(k+\frac{\Delta k}{2}\right)^2\right) = 4\left(\sigma^2+\frac{\Delta k^2}{4}\right). \end{equation}
\section{Schemes with detectors that do not resolve the transverse momenta} \label{app:NR}
In this Appendix we will evaluate, for the schemes that do not resolve the transverse momenta of the photons, in this order: the probability distribution, the saturation of the quantum Cramér-Rao bound for $\Delta x\simeq 0$, the analytical expression of the Fisher information with Gaussian wavepackets, and the first order expansion in $\Delta x$ of $\frac{F^\mathrm{NR}_{\Delta x}}{F_{\Delta x}}$. Since the precisions achieved when employing number-resolving detectors differ from the precision achieved with detectors that do not resolve the number of photons, we will analyze them separately.
\subsection{Number-resolving detectors} \subsubsection{Probability in Eq.~\eqref{eq:ProbNR}} Assuming initially that the detectors are photon-number resolving, the probability $P_\gamma^{\mathrm{NR}}(X;\Delta x)$ can be straightforwardly obtained by integrating the probability $P_\gamma( k',X;\Delta x)$ over all the observable transverse momenta $ k'$ \begin{align} P_\gamma^{\mathrm{NR}}(X;\Delta x)&= \frac{1}{4}(1-\gamma)^2\int\dd k'\ \left(\abs{f\left( k'- \frac{\Delta k}{2}\right)}^2+\abs{f\left( -\frac{\Delta k}{2}- k'\right)}^2\right)(1+\alpha(X)\cos(2k'\Delta x))\notag\\ &=\frac{1}{2}(1-\gamma)^2\int\dd k'\ \abs{f\left( k'-\frac{\Delta k}{2}\right)}^2(1+\alpha(X)\cos(2 k'\Delta x))\notag\\ &=\frac{1}{2}(1-\gamma)^2\left(1+\alpha(X)\Re\left[\e^{\ii\Delta k\Delta x}\int\dd k'\ \abs{f\left( k'\right)}^2\e^{2\ii k'\Delta x}\right]\right), \label{eq:ProbNRApp} \end{align} as shown in Eq.~\eqref{eq:ProbNR} in the main text, while $P_0=\gamma^2$ in Eq.~\eqref{eq:P0} and $\int\dd k'\ P_i(k')=\gamma(1-\gamma)$, for $i=1,2$ in Eq.~\eqref{eq:P1} are independent of $\Delta x$. \subsubsection{$F^{\mathrm{NR}}_{\Delta x}/F_{\Delta x}$ in the regime $\Delta x\ll 1/\Delta k$} We now evaluate the expression of the Fisher information $F^{\mathrm{NR}}_{\Delta x}$ for non-resolving measurements in the limit of $\Delta x\simeq 0$, and show that it saturates the quantum Fisher information $H_{\Delta x}$ in Eq.~\eqref{eq:FIs}. To do so, we first evaluate the derivatives \begin{align} \partial P^\mathrm{NR}_\gamma(X;\Delta x) &= -\frac{(1-\gamma)^2}{2}\alpha(X)\Im[\Delta k\ \e^{\ii\Delta k\Delta x}\int\dd k'\ \abs{f(k')}^2\e^{2\ii k'\Delta x}+2\e^{\ii\Delta k\Delta x}\int\dd k'\ k'\abs{f(k')}^2\e^{2\ii k'\Delta x}]\notag\\ &=-\frac{(1-\gamma)^2}{2}\alpha(X)\Im\Bigl[\Delta k \left(1+\ii\Delta k\Delta x-\frac{\Delta k^2\Delta x^2}{2}\right)\int\dd k'\ \abs{f(k')}^2(1+\ii 2k'\Delta x-2{k'}^2\Delta x^2)\notag\\ &\quad+2\left(1+\ii\Delta k\Delta x-\frac{\Delta k^2\Delta x^2}{2}\right)\int\dd k'\ k'\abs{f(k')}^2(1+\ii 2k'\Delta x-2{k'}^2\Delta x^2)\Big]+\O(\Delta x^3), \label{eq:IntermDeriv} \end{align} where $\O(\Delta x^d)$, $d\in\mathbb{N}$, represents a term of order of $\Delta x^d$ or higher which can be neglected for $\Delta x\simeq 0$. Once we perform the substitutions $\int\dd k'\ k'\abs{f(k')}^2 =0 $, $\int\dd k'\ k'^2\abs{f(k')}^2 =\sigma^2$, absorb the remaining terms of order $\Delta x^3$ and higher into $\O(\Delta x^3)$, and cancel the purely real terms inside the function $\Im[\cdot]$, Eq.~\eqref{eq:IntermDeriv} reduces to \begin{equation} \partial P^\mathrm{NR}_\gamma(X;\Delta x) = -\frac{(1-\gamma)^2}{2}\alpha(X)\Delta x(\Delta k^2 + 4\sigma^2) + \O(\Delta x^3). \label{eq:AsymDer} \end{equation} With similar steps, we can rewrite Eq.~\eqref{eq:ProbNRApp} as \begin{equation} P^\mathrm{NR}_\gamma(X;\Delta x) = \frac{(1-\gamma)^2}{2}\left(1+\alpha(X) - \frac{1}{2}\Delta x^2\alpha(X)(\Delta k^2+4\sigma^2)\right) +\O(\Delta x^4), \label{eq:AsymP} \end{equation} so that \begin{align} F^\mathrm{NR}_{\Delta x} &= \sum_{X=A,B}\frac{(\partial P^\mathrm{NR}_\gamma(X;\Delta x))^2}{P^\mathrm{NR}_\gamma(X;\Delta x)} \notag\\ &=\left(\frac{1}{4}\Delta x^2(\Delta k^2+4\sigma^2)^2+\O(\Delta x^4)\right)\left(\frac{(1-\gamma)^2}{1-\frac{1}{4}\Delta x^2(\Delta k^2+4\sigma^2)+\O(\Delta x^4)}+\frac{(1-\gamma)^2}{\frac{1}{4}\Delta x^2(\Delta k^2+4\sigma^2)+\O(\Delta x^4)}\right)\notag\\ &= (1-\gamma)^2\left(\frac{1}{4}\Delta x^2(\Delta k^2+4\sigma^2)^2+\O(\Delta x^4)\right)\left(1+\O(\Delta x^2)+\frac{1}{\frac{1}{4}\Delta x^2(\Delta k^2+4\sigma^2)}(1+\O(\Delta x^2))\right)\notag\\ &=(1-\gamma)^2\frac{\frac{1}{4}\Delta x^2(\Delta k^2+4\sigma^2)^2}{\frac{1}{4}\Delta x^2(\Delta k^2+4\sigma^2)}+\O(\Delta x^2)\simeq (1-\gamma)^2(\Delta k^2+4\sigma^2)=F_{\Delta x} \label{eq:Asym0} \end{align} where we are neglecting terms of order $\O(\Delta x^2)$. Notice that in the last step we removed a discontinuity in $\Delta x=0$. We can see that $F^\mathrm{NR}_{\Delta x}$ coincides with the quantum Fisher information when neglecting terms of order $\O(\Delta x)$, i.e. for $\Delta x\simeq 0$, and lossless detection $\gamma=0$. \subsubsection{Fisher information in Eq.~\eqref{eq:FINR} for Gaussian wavepackets and small $\Delta x$ expansion in Eq.~\eqref{eq:FINRsmall}} For a Gaussian distribution $\abs{f( k)}^2$, the probability $P_\gamma^{\mathrm{NR}}(X;\Delta x)$ specializes to \begin{equation} P_\gamma^{\mathrm{NR}}(X;\Delta x) = \frac{(1-\gamma)^2}{2}\left(1+\alpha(X)\exp(-2\sigma^2\Delta x^2)\cos(\Delta k\Delta x)\right). \end{equation} Applying the definition of the Fisher information for the non-resolving scheme, we can easily evaluate \begin{equation} F^\mathrm{NR}_{\Delta x} = \mathbb{E}\left[(\partial\log P_\gamma^\mathrm{NR}(X;\Delta x))^2\right]= (1-\gamma)^2 \e^{-4\sigma^2\Delta x^2}\frac{(4\sigma^2\Delta x\cos(\Delta k\Delta x)+\Delta k\sin(\Delta k\Delta x))^2}{1-\cos^2(\Delta k\Delta x)\e^{-4\sigma^2\Delta x^2}}, \label{eq:FINRApp} \end{equation} as shown in Eq.~\eqref{eq:FINR} in the main text. Expanding Eq.\eqref{eq:FINRApp} in series for $\Delta x=0$ and normalizing over the Fisher information of the resolving scheme $F_{\Delta x}$, we can check how quickly $F^\mathrm{NR}_{\Delta x}$ decreases from its maximum in $\Delta x=0$ for Gaussian distribution, namely \begin{equation} \frac{F^\mathrm{NR}_{\Delta x}}{F_{\Delta x}}=1 -4\frac{\Delta k^2+2\sigma^2}{\Delta k^2+4\sigma^2}\sigma^2\Delta x^2+\O(\Delta x^4)\simeq 1 -4 \sigma^2\Delta x^2+\O(\Delta x^4), \end{equation} where the last approximation is justified since we are assuming $\Delta k\gg\sigma$.

\subsection{On-off detectors} \subsubsection{Probability in Eq.~\eqref{eq:ProbNNR}} We observe that, in the presence of losses, if the detectors cannot distinguish the number of photons that arrived, it becomes impossible to discriminate between the event of a loss of a single photon and a bunching event, as the interested detector only clicks once in either case. Recalling the expressions of $P_i(k')$ and $P_0$ in \cref{eq:P0,eq:P1}, we can evaluate the probabilities of the three possible outcomes of double, single, and no click observations as, respectively \begin{align} \mathcal{P}^{\mathrm{NR}}_{\gamma,2}(\Delta x) &= P^\mathrm{NR}_\gamma(A;\Delta x)=\frac{(1-\gamma)^2}{2}\left(1-\Re\left[\e^{\ii\Delta k\Delta x}\int\dd k'\ \abs{f\left( k'\right)}^2\e^{2\ii k'\Delta x}\right]\right),\\ \mathcal{P}^{\mathrm{NR}}_{\gamma,1}(\Delta x) &= P^\mathrm{NR}_\gamma(B;\Delta x)+ \int\dd k' \sum_{i=1,2} P_i(k')=\frac{(1-\gamma)^2}{2}\left(\frac{1+3\gamma}{1-\gamma}+\Re\left[\e^{\ii\Delta k\Delta x}\int\dd k'\ \abs{f\left( k'\right)}^2\e^{2\ii k'\Delta x}\right]\right),\\ \mathcal{P}^{\mathrm{NR}}_{\gamma,0}&=P_0 =\gamma^2. \end{align}
\subsubsection{$\mathcal{F}^{\mathrm{NR}}_{\Delta x}/F_{\Delta x}$ in the regime $\Delta x\ll 1/\Delta k$} We can easily check that the behavior for $\Delta x\rightarrow 0$ of the Fisher information $\mathcal{F}^{\mathrm{NR}}_{\Delta x}$ that can be evaluated from these probabilities is identical to Eq.~\eqref{eq:Asym0}. Indeed, the derivatives $\partial \mathcal{P}^{\mathrm{NR}}_{\gamma,i}$ remain the same as $\partial P^{\mathrm{NR}}_{\gamma}(X;\Delta x)$ in Eq.~\eqref{eq:IntermDeriv}, so their expansion for small $\Delta x$ in Eq.~\eqref{eq:AsymDer}, as well as the one of $\mathcal{P}^{\mathrm{NR}}_{\gamma,2} =P^\mathrm{NR}_{\gamma}(A,\Delta x)$ in Eq.~\eqref{eq:AsymP} remain unchanged, while \begin{align} \mathcal{P}^{\mathrm{NR}}_{\gamma,1}(\Delta x) = \frac{(1-\gamma)^2}{2}\left(2\frac{1+\gamma}{1-\gamma} - \frac{1}{2}\Delta x^2(\Delta k^2+4\sigma^2)\right) + \O(\Delta x^4), \end{align} so that \begin{align} \mathcal{F}^{\mathrm{NR}}_{\Delta x} &= \sum_{i=1,2}\frac{(\partial \mathcal{P}^{\mathrm{NR}}_{\gamma,i}(\Delta x))^2}{\mathcal{P}^{\mathrm{NR}}_{\gamma,i}(\Delta x)} \notag\\ &=\left(\frac{1}{4}\Delta x^2(\Delta k^2+4\sigma^2)^2+\O(\Delta x^4)\right)\left(\frac{(1-\gamma)^2}{\frac{1+\gamma}{1-\gamma}-\frac{1}{4}\Delta x^2(\Delta k^2+4\sigma^2)+\O(\Delta x^4)}+\frac{(1-\gamma)^2}{\frac{1}{4}\Delta x^2(\Delta k^2+4\sigma^2)+\O(\Delta x^4)}\right)\notag\\ &= (1-\gamma)^2\left(\frac{1}{4}\Delta x^2(\Delta k^2+4\sigma^2)^2+\O(\Delta x^4)\right)\left(\frac{1-\gamma}{1+\gamma}+\O(\Delta x^2)+\frac{1}{\frac{1}{4}\Delta x^2(\Delta k^2+4\sigma^2)}(1+\O(\Delta x^2))\right)\notag\\ &=(1-\gamma)^2\frac{\frac{1}{4}\Delta x^2(\Delta k^2+4\sigma^2)^2}{\frac{1}{4}\Delta x^2(\Delta k^2+4\sigma^2)}+\O(\Delta x^2)\simeq (1-\gamma)^2(\Delta k^2+4\sigma^2) = F_{\Delta x} \end{align} \subsubsection{Fisher information in Eq.~\eqref{eq:FINNR} for Gaussian wavepackets and small $\Delta x$ expansion in Eq.~\eqref{eq:FINNRsmall}} Finally, for Gaussian wavepackets, we have \begin{equation} \mathcal{F}^{\mathrm{NR}}_{\Delta x} = \mathbb{E}\left[(\partial\log P_{\gamma,i}^\mathrm{NNR}(\Delta x))^2\right]= \frac{(1-\gamma^2) \e^{-4\sigma^2\Delta x^2}(4\sigma^2\Delta x\cos(\Delta k\Delta x)+\Delta k\sin(\Delta k\Delta x))^2}{(1-\exp(-2\sigma^2\Delta x^2)\cos(\Delta k\Delta x))(\frac{1+3\gamma}{1-\gamma}+\exp(-2\sigma^2\Delta x^2)\cos(\Delta k\Delta x))}. \label{eq:appFINNR} \end{equation} Once again, we can expand Eq.~\eqref{eq:appFINNR} in series for $\Delta x=0$ and normalize it over $F_{\Delta x}$ and obtain \begin{equation} \frac{\mathcal{F}^{\mathrm{NR}}_{\Delta x}}{F_{\Delta x}}= 1 - \frac{1}{2(1+\gamma)}\frac{\gamma\Delta k^4+8\sigma^2(1+2\gamma)(\Delta k^2+2\sigma^2)}{\Delta k^2+4\sigma^2}\Delta x^2+\O(\Delta x^4)= 1 - \frac{\gamma}{2(1+\gamma)}\Delta k^2\Delta x^2+ \O(\sigma^2\Delta x^2)+\O(\Delta x^4), \end{equation} where the terms of order $\O(\sigma^2\Delta x^2)$ can be neglected for $\Delta k\gg \sigma$ and $\gamma\neq0$.
\twocolumngrid
\bibliography{references}

\begin{thebibliography}{31}%
\makeatletter
\providecommand \@ifxundefined [1]{%
 \@ifx{#1\undefined}
}%
\providecommand \@ifnum [1]{%
 \ifnum #1\expandafter \@firstoftwo
 \else \expandafter \@secondoftwo
 \fi
}%
\providecommand \@ifx [1]{%
 \ifx #1\expandafter \@firstoftwo
 \else \expandafter \@secondoftwo
 \fi
}%
\providecommand \natexlab [1]{#1}%
\providecommand \enquote  [1]{``#1''}%
\providecommand \bibnamefont  [1]{#1}%
\providecommand \bibfnamefont [1]{#1}%
\providecommand \citenamefont [1]{#1}%
\providecommand \href@noop [0]{\@secondoftwo}%
\providecommand \href [0]{\begingroup \@sanitize@url \@href}%
\providecommand \@href[1]{\@@startlink{#1}\@@href}%
\providecommand \@@href[1]{\endgroup#1\@@endlink}%
\providecommand \@sanitize@url [0]{\catcode `\\12\catcode `\$12\catcode
  `\&12\catcode `\#12\catcode `\^12\catcode `\_12\catcode `\%12\relax}%
\providecommand \@@startlink[1]{}%
\providecommand \@@endlink[0]{}%
\providecommand \url  [0]{\begingroup\@sanitize@url \@url }%
\providecommand \@url [1]{\endgroup\@href {#1}{\urlprefix }}%
\providecommand \urlprefix  [0]{URL }%
\providecommand \Eprint [0]{\href }%
\providecommand \doibase [0]{https://doi.org/}%
\providecommand \selectlanguage [0]{\@gobble}%
\providecommand \bibinfo  [0]{\@secondoftwo}%
\providecommand \bibfield  [0]{\@secondoftwo}%
\providecommand \translation [1]{[#1]}%
\providecommand \BibitemOpen [0]{}%
\providecommand \bibitemStop [0]{}%
\providecommand \bibitemNoStop [0]{.\EOS\space}%
\providecommand \EOS [0]{\spacefactor3000\relax}%
\providecommand \BibitemShut  [1]{\csname bibitem#1\endcsname}%
\let\auto@bib@innerbib\@empty
\bibitem [{\citenamefont {Hong}\ \emph {et~al.}(1987)\citenamefont {Hong},
  \citenamefont {Ou},\ and\ \citenamefont {Mandel}}]{Hong1987}%
  \BibitemOpen
  \bibfield  {author} {\bibinfo {author} {\bibfnamefont {C.~K.}\ \bibnamefont
  {Hong}}, \bibinfo {author} {\bibfnamefont {Z.~Y.}\ \bibnamefont {Ou}},\ and\
  \bibinfo {author} {\bibfnamefont {L.}~\bibnamefont {Mandel}},\ }\bibfield
  {title} {\bibinfo {title} {Measurement of subpicosecond time intervals
  between two photons by interference},\ }\href
  {https://doi.org/10.1103/PhysRevLett.59.2044} {\bibfield  {journal} {\bibinfo
   {journal} {Phys. Rev. Lett.}\ }\textbf {\bibinfo {volume} {59}},\ \bibinfo
  {pages} {2044} (\bibinfo {year} {1987})}\BibitemShut {NoStop}%
\bibitem [{\citenamefont {Shih}\ and\ \citenamefont {Alley}(1988)}]{Shih1988}%
  \BibitemOpen
  \bibfield  {author} {\bibinfo {author} {\bibfnamefont {Y.~H.}\ \bibnamefont
  {Shih}}\ and\ \bibinfo {author} {\bibfnamefont {C.~O.}\ \bibnamefont
  {Alley}},\ }\bibfield  {title} {\bibinfo {title} {New type of
  einstein-podolsky-rosen-bohm experiment using pairs of light quanta produced
  by optical parametric down conversion},\ }\href
  {https://doi.org/10.1103/PhysRevLett.61.2921} {\bibfield  {journal} {\bibinfo
   {journal} {Phys. Rev. Lett.}\ }\textbf {\bibinfo {volume} {61}},\ \bibinfo
  {pages} {2921} (\bibinfo {year} {1988})}\BibitemShut {NoStop}%
\bibitem [{\citenamefont {Bouchard}\ \emph {et~al.}(2021)\citenamefont
  {Bouchard}, \citenamefont {Sit}, \citenamefont {Zhang}, \citenamefont
  {Fickler}, \citenamefont {Miatto}, \citenamefont {Yao}, \citenamefont
  {Sciarrino},\ and\ \citenamefont {Karimi}}]{Bouchard2021}%
  \BibitemOpen
  \bibfield  {author} {\bibinfo {author} {\bibfnamefont {F.}~\bibnamefont
  {Bouchard}}, \bibinfo {author} {\bibfnamefont {A.}~\bibnamefont {Sit}},
  \bibinfo {author} {\bibfnamefont {Y.}~\bibnamefont {Zhang}}, \bibinfo
  {author} {\bibfnamefont {R.}~\bibnamefont {Fickler}}, \bibinfo {author}
  {\bibfnamefont {F.~M.}\ \bibnamefont {Miatto}}, \bibinfo {author}
  {\bibfnamefont {Y.}~\bibnamefont {Yao}}, \bibinfo {author} {\bibfnamefont
  {F.}~\bibnamefont {Sciarrino}},\ and\ \bibinfo {author} {\bibfnamefont
  {E.}~\bibnamefont {Karimi}},\ }\bibfield  {title} {\bibinfo {title}
  {Two-photon interference: the hong-ou-mandel effect},\ }\href
  {https://doi.org/10.1088/1361-6633/abcd7a} {\bibfield  {journal} {\bibinfo
  {journal} {Reports on Progress in Physics}\ }\textbf {\bibinfo {volume}
  {84}},\ \bibinfo {pages} {012402} (\bibinfo {year} {2021})}\BibitemShut
  {NoStop}%
\bibitem [{\citenamefont {Legero}\ \emph {et~al.}(2003)\citenamefont {Legero},
  \citenamefont {Wilk}, \citenamefont {Kuhn},\ and\ \citenamefont
  {Rempe}}]{Legero2003}%
  \BibitemOpen
  \bibfield  {author} {\bibinfo {author} {\bibfnamefont {T.}~\bibnamefont
  {Legero}}, \bibinfo {author} {\bibfnamefont {T.}~\bibnamefont {Wilk}},
  \bibinfo {author} {\bibfnamefont {A.}~\bibnamefont {Kuhn}},\ and\ \bibinfo
  {author} {\bibfnamefont {G.}~\bibnamefont {Rempe}},\ }\bibfield  {title}
  {\bibinfo {title} {Time-resolved two-photon quantum interference},\ }\href
  {https://doi.org/10.1007/s00340-003-1337-x} {\bibfield  {journal} {\bibinfo
  {journal} {Applied Physics B}\ }\textbf {\bibinfo {volume} {77}},\ \bibinfo
  {pages} {797} (\bibinfo {year} {2003})}\BibitemShut {NoStop}%
\bibitem [{\citenamefont {Legero}\ \emph {et~al.}(2004)\citenamefont {Legero},
  \citenamefont {Wilk}, \citenamefont {Hennrich}, \citenamefont {Rempe},\ and\
  \citenamefont {Kuhn}}]{Legero2004}%
  \BibitemOpen
  \bibfield  {author} {\bibinfo {author} {\bibfnamefont {T.}~\bibnamefont
  {Legero}}, \bibinfo {author} {\bibfnamefont {T.}~\bibnamefont {Wilk}},
  \bibinfo {author} {\bibfnamefont {M.}~\bibnamefont {Hennrich}}, \bibinfo
  {author} {\bibfnamefont {G.}~\bibnamefont {Rempe}},\ and\ \bibinfo {author}
  {\bibfnamefont {A.}~\bibnamefont {Kuhn}},\ }\bibfield  {title} {\bibinfo
  {title} {Quantum beat of two single photons},\ }\href
  {https://doi.org/10.1103/PhysRevLett.93.070503} {\bibfield  {journal}
  {\bibinfo  {journal} {Phys. Rev. Lett.}\ }\textbf {\bibinfo {volume} {93}},\
  \bibinfo {pages} {070503} (\bibinfo {year} {2004})}\BibitemShut {NoStop}%
\bibitem [{\citenamefont {Chen}\ \emph {et~al.}(2023)\citenamefont {Chen},
  \citenamefont {Chen},\ and\ \citenamefont {Chen}}]{Chen2023}%
  \BibitemOpen
  \bibfield  {author} {\bibinfo {author} {\bibfnamefont {C.}~\bibnamefont
  {Chen}}, \bibinfo {author} {\bibfnamefont {Y.}~\bibnamefont {Chen}},\ and\
  \bibinfo {author} {\bibfnamefont {L.}~\bibnamefont {Chen}},\ }\bibfield
  {title} {\bibinfo {title} {Spectrally resolved hong-ou-mandel interferometry
  with discrete color entanglement},\ }\href
  {https://doi.org/10.1103/PhysRevApplied.19.054092} {\bibfield  {journal}
  {\bibinfo  {journal} {Phys. Rev. Appl.}\ }\textbf {\bibinfo {volume} {19}},\
  \bibinfo {pages} {054092} (\bibinfo {year} {2023})}\BibitemShut {NoStop}%
\bibitem [{\citenamefont {Lyons}\ \emph {et~al.}(2018)\citenamefont {Lyons},
  \citenamefont {Knee}, \citenamefont {Bolduc}, \citenamefont {Roger},
  \citenamefont {Leach}, \citenamefont {Gauger},\ and\ \citenamefont
  {Faccio}}]{Lyons2018}%
  \BibitemOpen
  \bibfield  {author} {\bibinfo {author} {\bibfnamefont {A.}~\bibnamefont
  {Lyons}}, \bibinfo {author} {\bibfnamefont {G.~C.}\ \bibnamefont {Knee}},
  \bibinfo {author} {\bibfnamefont {E.}~\bibnamefont {Bolduc}}, \bibinfo
  {author} {\bibfnamefont {T.}~\bibnamefont {Roger}}, \bibinfo {author}
  {\bibfnamefont {J.}~\bibnamefont {Leach}}, \bibinfo {author} {\bibfnamefont
  {E.~M.}\ \bibnamefont {Gauger}},\ and\ \bibinfo {author} {\bibfnamefont
  {D.}~\bibnamefont {Faccio}},\ }\bibfield  {title} {\bibinfo {title}
  {Attosecond-resolution hong-ou-mandel interferometry},\ }\href
  {https://advances.sciencemag.org/content/4/5/eaap9416} {\bibfield  {journal}
  {\bibinfo  {journal} {Science Advances}\ }\textbf {\bibinfo {volume} {4}},\
  \bibinfo {pages} {5} (\bibinfo {year} {2018})}\BibitemShut {NoStop}%
\bibitem [{\citenamefont {Scott}\ \emph {et~al.}(2020)\citenamefont {Scott},
  \citenamefont {Branford}, \citenamefont {Westerberg}, \citenamefont {Leach},\
  and\ \citenamefont {Gauger}}]{Scott2020}%
  \BibitemOpen
  \bibfield  {author} {\bibinfo {author} {\bibfnamefont {H.}~\bibnamefont
  {Scott}}, \bibinfo {author} {\bibfnamefont {D.}~\bibnamefont {Branford}},
  \bibinfo {author} {\bibfnamefont {N.}~\bibnamefont {Westerberg}}, \bibinfo
  {author} {\bibfnamefont {J.}~\bibnamefont {Leach}},\ and\ \bibinfo {author}
  {\bibfnamefont {E.~M.}\ \bibnamefont {Gauger}},\ }\bibfield  {title}
  {\bibinfo {title} {Beyond coincidence in hong-ou-mandel interferometry},\
  }\href {https://doi.org/10.1103/PhysRevA.102.033714} {\bibfield  {journal}
  {\bibinfo  {journal} {Phys. Rev. A}\ }\textbf {\bibinfo {volume} {102}},\
  \bibinfo {pages} {033714} (\bibinfo {year} {2020})}\BibitemShut {NoStop}%
\bibitem [{\citenamefont {Fabre}\ and\ \citenamefont
  {Felicetti}(2021)}]{Fabre2021}%
  \BibitemOpen
  \bibfield  {author} {\bibinfo {author} {\bibfnamefont {N.}~\bibnamefont
  {Fabre}}\ and\ \bibinfo {author} {\bibfnamefont {S.}~\bibnamefont
  {Felicetti}},\ }\bibfield  {title} {\bibinfo {title} {Parameter estimation of
  time and frequency shifts with generalized hong-ou-mandel interferometry},\
  }\href {https://doi.org/10.1103/PhysRevA.104.022208} {\bibfield  {journal}
  {\bibinfo  {journal} {Phys. Rev. A}\ }\textbf {\bibinfo {volume} {104}},\
  \bibinfo {pages} {022208} (\bibinfo {year} {2021})}\BibitemShut {NoStop}%
\bibitem [{\citenamefont {Harnchaiwat}\ \emph {et~al.}(2020)\citenamefont
  {Harnchaiwat}, \citenamefont {Zhu}, \citenamefont {Westerberg}, \citenamefont
  {Gauger},\ and\ \citenamefont {Leach}}]{Harnchaiwat2020}%
  \BibitemOpen
  \bibfield  {author} {\bibinfo {author} {\bibfnamefont {N.}~\bibnamefont
  {Harnchaiwat}}, \bibinfo {author} {\bibfnamefont {F.}~\bibnamefont {Zhu}},
  \bibinfo {author} {\bibfnamefont {N.}~\bibnamefont {Westerberg}}, \bibinfo
  {author} {\bibfnamefont {E.}~\bibnamefont {Gauger}},\ and\ \bibinfo {author}
  {\bibfnamefont {J.}~\bibnamefont {Leach}},\ }\bibfield  {title} {\bibinfo
  {title} {Tracking the polarisation state of light via hong-ou-mandel
  interferometry},\ }\href {https://doi.org/10.1364/OE.382622} {\bibfield
  {journal} {\bibinfo  {journal} {Opt. Express}\ }\textbf {\bibinfo {volume}
  {28}},\ \bibinfo {pages} {2210} (\bibinfo {year} {2020})}\BibitemShut
  {NoStop}%
\bibitem [{\citenamefont {Sgobba}\ \emph {et~al.}(2023)\citenamefont {Sgobba},
  \citenamefont {Pallotti}, \citenamefont {Elefante}, \citenamefont
  {Dello~Russo}, \citenamefont {Dequal}, \citenamefont {Siciliani~de Cumis},\
  and\ \citenamefont {Santamaria~Amato}}]{Sgobba2023}%
  \BibitemOpen
  \bibfield  {author} {\bibinfo {author} {\bibfnamefont {F.}~\bibnamefont
  {Sgobba}}, \bibinfo {author} {\bibfnamefont {D.~K.}\ \bibnamefont
  {Pallotti}}, \bibinfo {author} {\bibfnamefont {A.}~\bibnamefont {Elefante}},
  \bibinfo {author} {\bibfnamefont {S.}~\bibnamefont {Dello~Russo}}, \bibinfo
  {author} {\bibfnamefont {D.}~\bibnamefont {Dequal}}, \bibinfo {author}
  {\bibfnamefont {M.}~\bibnamefont {Siciliani~de Cumis}},\ and\ \bibinfo
  {author} {\bibfnamefont {L.}~\bibnamefont {Santamaria~Amato}},\ }\bibfield
  {title} {\bibinfo {title} {Optimal measurement of telecom wavelength single
  photon polarisation via hong-ou-mandel interferometry},\ }\bibfield
  {journal} {\bibinfo  {journal} {Photonics}\ }\textbf {\bibinfo {volume}
  {10}},\ \href {https://doi.org/10.3390/photonics10010072}
  {10.3390/photonics10010072} (\bibinfo {year} {2023})\BibitemShut {NoStop}%
\bibitem [{\citenamefont {Abouraddy}\ \emph {et~al.}(2002)\citenamefont
  {Abouraddy}, \citenamefont {Nasr}, \citenamefont {Saleh}, \citenamefont
  {Sergienko},\ and\ \citenamefont {Teich}}]{Abouraddy2002}%
  \BibitemOpen
  \bibfield  {author} {\bibinfo {author} {\bibfnamefont {A.~F.}\ \bibnamefont
  {Abouraddy}}, \bibinfo {author} {\bibfnamefont {M.~B.}\ \bibnamefont {Nasr}},
  \bibinfo {author} {\bibfnamefont {B.~E.~A.}\ \bibnamefont {Saleh}}, \bibinfo
  {author} {\bibfnamefont {A.~V.}\ \bibnamefont {Sergienko}},\ and\ \bibinfo
  {author} {\bibfnamefont {M.~C.}\ \bibnamefont {Teich}},\ }\bibfield  {title}
  {\bibinfo {title} {Quantum-optical coherence tomography with dispersion
  cancellation},\ }\href {https://doi.org/10.1103/PhysRevA.65.053817}
  {\bibfield  {journal} {\bibinfo  {journal} {Phys. Rev. A}\ }\textbf {\bibinfo
  {volume} {65}},\ \bibinfo {pages} {053817} (\bibinfo {year}
  {2002})}\BibitemShut {NoStop}%
\bibitem [{\citenamefont {Nasr}\ \emph {et~al.}(2009)\citenamefont {Nasr},
  \citenamefont {Goode}, \citenamefont {Nguyen}, \citenamefont {Rong},
  \citenamefont {Yang}, \citenamefont {Reinhard}, \citenamefont {Saleh},\ and\
  \citenamefont {Teich}}]{Nasr2009}%
  \BibitemOpen
  \bibfield  {author} {\bibinfo {author} {\bibfnamefont {M.~B.}\ \bibnamefont
  {Nasr}}, \bibinfo {author} {\bibfnamefont {D.~P.}\ \bibnamefont {Goode}},
  \bibinfo {author} {\bibfnamefont {N.}~\bibnamefont {Nguyen}}, \bibinfo
  {author} {\bibfnamefont {G.}~\bibnamefont {Rong}}, \bibinfo {author}
  {\bibfnamefont {L.}~\bibnamefont {Yang}}, \bibinfo {author} {\bibfnamefont
  {B.~M.}\ \bibnamefont {Reinhard}}, \bibinfo {author} {\bibfnamefont {B.~E.}\
  \bibnamefont {Saleh}},\ and\ \bibinfo {author} {\bibfnamefont {M.~C.}\
  \bibnamefont {Teich}},\ }\bibfield  {title} {\bibinfo {title} {Quantum
  optical coherence tomography of a biological sample},\ }\href
  {https://doi.org/https://doi.org/10.1016/j.optcom.2008.11.061} {\bibfield
  {journal} {\bibinfo  {journal} {Optics Communications}\ }\textbf {\bibinfo
  {volume} {282}},\ \bibinfo {pages} {1154} (\bibinfo {year}
  {2009})}\BibitemShut {NoStop}%
\bibitem [{\citenamefont {Yepiz-Graciano}\ \emph {et~al.}(2020)\citenamefont
  {Yepiz-Graciano}, \citenamefont {Mart\'{i}nez}, \citenamefont {Lopez-Mago},
  \citenamefont {Cruz-Ramirez},\ and\ \citenamefont
  {U'Ren}}]{Yepiz-Graciano2020}%
  \BibitemOpen
  \bibfield  {author} {\bibinfo {author} {\bibfnamefont {P.}~\bibnamefont
  {Yepiz-Graciano}}, \bibinfo {author} {\bibfnamefont {A.~M.~A.}\ \bibnamefont
  {Mart\'{i}nez}}, \bibinfo {author} {\bibfnamefont {D.}~\bibnamefont
  {Lopez-Mago}}, \bibinfo {author} {\bibfnamefont {H.}~\bibnamefont
  {Cruz-Ramirez}},\ and\ \bibinfo {author} {\bibfnamefont {A.~B.}\ \bibnamefont
  {U'Ren}},\ }\bibfield  {title} {\bibinfo {title} {Spectrally resolved
  hong-ou-mandel interferometry for quantum-optical coherence tomography},\
  }\href {https://doi.org/10.1364/PRJ.388693} {\bibfield  {journal} {\bibinfo
  {journal} {Photon. Res.}\ }\textbf {\bibinfo {volume} {8}},\ \bibinfo {pages}
  {1023} (\bibinfo {year} {2020})}\BibitemShut {NoStop}%
\bibitem [{\citenamefont {Hayama}\ \emph {et~al.}(2022)\citenamefont {Hayama},
  \citenamefont {Cao}, \citenamefont {Okamoto}, \citenamefont {Suezawa},
  \citenamefont {Okano},\ and\ \citenamefont {Takeuchi}}]{Hayama2022}%
  \BibitemOpen
  \bibfield  {author} {\bibinfo {author} {\bibfnamefont {K.}~\bibnamefont
  {Hayama}}, \bibinfo {author} {\bibfnamefont {B.}~\bibnamefont {Cao}},
  \bibinfo {author} {\bibfnamefont {R.}~\bibnamefont {Okamoto}}, \bibinfo
  {author} {\bibfnamefont {S.}~\bibnamefont {Suezawa}}, \bibinfo {author}
  {\bibfnamefont {M.}~\bibnamefont {Okano}},\ and\ \bibinfo {author}
  {\bibfnamefont {S.}~\bibnamefont {Takeuchi}},\ }\bibfield  {title} {\bibinfo
  {title} {High-depth-resolution imaging of dispersive samples using quantum
  optical coherence tomography},\ }\href {https://doi.org/10.1364/OL.469874}
  {\bibfield  {journal} {\bibinfo  {journal} {Opt. Lett.}\ }\textbf {\bibinfo
  {volume} {47}},\ \bibinfo {pages} {4949} (\bibinfo {year}
  {2022})}\BibitemShut {NoStop}%
\bibitem [{\citenamefont {Lyons}\ \emph {et~al.}(2023)\citenamefont {Lyons},
  \citenamefont {Zickus}, \citenamefont {{\'A}lvarez-Mendoza}, \citenamefont
  {Triggiani}, \citenamefont {Tamma}, \citenamefont {Westerberg}, \citenamefont
  {Tassieri},\ and\ \citenamefont {Faccio}}]{Lyons2023}%
  \BibitemOpen
  \bibfield  {author} {\bibinfo {author} {\bibfnamefont {A.}~\bibnamefont
  {Lyons}}, \bibinfo {author} {\bibfnamefont {V.}~\bibnamefont {Zickus}},
  \bibinfo {author} {\bibfnamefont {R.}~\bibnamefont {{\'A}lvarez-Mendoza}},
  \bibinfo {author} {\bibfnamefont {D.}~\bibnamefont {Triggiani}}, \bibinfo
  {author} {\bibfnamefont {V.}~\bibnamefont {Tamma}}, \bibinfo {author}
  {\bibfnamefont {N.}~\bibnamefont {Westerberg}}, \bibinfo {author}
  {\bibfnamefont {M.}~\bibnamefont {Tassieri}},\ and\ \bibinfo {author}
  {\bibfnamefont {D.}~\bibnamefont {Faccio}},\ }\bibfield  {title} {\bibinfo
  {title} {Fluorescence lifetime hong-ou-mandel sensing},\ }\href
  {https://doi.org/10.1038/s41467-023-43868-x} {\bibfield  {journal} {\bibinfo
  {journal} {Nature Communications}\ }\textbf {\bibinfo {volume} {14}},\
  \bibinfo {pages} {8005} (\bibinfo {year} {2023})}\BibitemShut {NoStop}%
\bibitem [{\citenamefont {Triggiani}\ and\ \citenamefont
  {Tamma}(2024)}]{Triggiani2024}%
  \BibitemOpen
  \bibfield  {author} {\bibinfo {author} {\bibfnamefont {D.}~\bibnamefont
  {Triggiani}}\ and\ \bibinfo {author} {\bibfnamefont {V.}~\bibnamefont
  {Tamma}},\ }\bibfield  {title} {\bibinfo {title} {Estimation with ultimate
  quantum precision of the transverse displacement between two photons via
  two-photon interference sampling measurements},\ }\href
  {https://doi.org/10.1103/PhysRevLett.132.180802} {\bibfield  {journal}
  {\bibinfo  {journal} {Phys. Rev. Lett.}\ }\textbf {\bibinfo {volume} {132}},\
  \bibinfo {pages} {180802} (\bibinfo {year} {2024})}\BibitemShut {NoStop}%
\bibitem [{\citenamefont {Chen}\ \emph {et~al.}(2019)\citenamefont {Chen},
  \citenamefont {Fink}, \citenamefont {Steinlechner}, \citenamefont {Torres},\
  and\ \citenamefont {Ursin}}]{Chen2019}%
  \BibitemOpen
  \bibfield  {author} {\bibinfo {author} {\bibfnamefont {Y.}~\bibnamefont
  {Chen}}, \bibinfo {author} {\bibfnamefont {M.}~\bibnamefont {Fink}}, \bibinfo
  {author} {\bibfnamefont {F.}~\bibnamefont {Steinlechner}}, \bibinfo {author}
  {\bibfnamefont {J.~P.}\ \bibnamefont {Torres}},\ and\ \bibinfo {author}
  {\bibfnamefont {R.}~\bibnamefont {Ursin}},\ }\bibfield  {title} {\bibinfo
  {title} {Hong-ou-mandel interferometry on a biphoton beat note},\ }\href
  {https://doi.org/10.1038/s41534-019-0161-z} {\bibfield  {journal} {\bibinfo
  {journal} {npj Quantum Information}\ }\textbf {\bibinfo {volume} {5}},\
  \bibinfo {pages} {43} (\bibinfo {year} {2019})}\BibitemShut {NoStop}%
\bibitem [{\citenamefont {Triggiani}\ \emph {et~al.}(2023)\citenamefont
  {Triggiani}, \citenamefont {Psaroudis},\ and\ \citenamefont
  {Tamma}}]{Triggiani2023}%
  \BibitemOpen
  \bibfield  {author} {\bibinfo {author} {\bibfnamefont {D.}~\bibnamefont
  {Triggiani}}, \bibinfo {author} {\bibfnamefont {G.}~\bibnamefont
  {Psaroudis}},\ and\ \bibinfo {author} {\bibfnamefont {V.}~\bibnamefont
  {Tamma}},\ }\bibfield  {title} {\bibinfo {title} {Ultimate quantum
  sensitivity in the estimation of the delay between two interfering photons
  through frequency-resolving sampling},\ }\href
  {https://doi.org/10.1103/PhysRevApplied.19.044068} {\bibfield  {journal}
  {\bibinfo  {journal} {Phys. Rev. Appl.}\ }\textbf {\bibinfo {volume} {19}},\
  \bibinfo {pages} {044068} (\bibinfo {year} {2023})}\BibitemShut {NoStop}%
\bibitem [{\citenamefont {Maggio}\ \emph {et~al.}(2024)\citenamefont {Maggio},
  \citenamefont {Triggiani}, \citenamefont {Facchi},\ and\ \citenamefont
  {Tamma}}]{Maggio2024}%
  \BibitemOpen
  \bibfield  {author} {\bibinfo {author} {\bibfnamefont {L.}~\bibnamefont
  {Maggio}}, \bibinfo {author} {\bibfnamefont {D.}~\bibnamefont {Triggiani}},
  \bibinfo {author} {\bibfnamefont {P.}~\bibnamefont {Facchi}},\ and\ \bibinfo
  {author} {\bibfnamefont {V.}~\bibnamefont {Tamma}},\ }\href@noop {} {\bibinfo
  {title} {Multi-parameter estimation of the state of two interfering photonic
  qubits}} (\bibinfo {year} {2024}),\ \Eprint
  {https://arxiv.org/abs/2405.12870} {arXiv:2405.12870} \BibitemShut {NoStop}%
\bibitem [{\citenamefont {Muratore}\ \emph {et~al.}(2024)\citenamefont
  {Muratore}, \citenamefont {Triggiani},\ and\ \citenamefont
  {Tamma}}]{Muratore2024}%
  \BibitemOpen
  \bibfield  {author} {\bibinfo {author} {\bibfnamefont {S.}~\bibnamefont
  {Muratore}}, \bibinfo {author} {\bibfnamefont {D.}~\bibnamefont
  {Triggiani}},\ and\ \bibinfo {author} {\bibfnamefont {V.}~\bibnamefont
  {Tamma}},\ }\href {https://arxiv.org/abs/2412.10057} {\bibinfo {title}
  {Superresolution imaging of two incoherent sources via two-photon
  interference sampling measurements in the transverse momenta}} (\bibinfo
  {year} {2024}),\ \Eprint {https://arxiv.org/abs/2412.10057} {arXiv:2412.10057
  [quant-ph]} \BibitemShut {NoStop}%
\bibitem [{\citenamefont {Salazar-Serrano}\ \emph {et~al.}(2015)\citenamefont
  {Salazar-Serrano}, \citenamefont {Valencia},\ and\ \citenamefont
  {Torres}}]{Serrano2015}%
  \BibitemOpen
  \bibfield  {author} {\bibinfo {author} {\bibfnamefont {L.~J.}\ \bibnamefont
  {Salazar-Serrano}}, \bibinfo {author} {\bibfnamefont {A.}~\bibnamefont
  {Valencia}},\ and\ \bibinfo {author} {\bibfnamefont {J.~P.}\ \bibnamefont
  {Torres}},\ }\bibfield  {title} {\bibinfo {title} {{Tunable beam
  displacer}},\ }\bibfield  {journal} {\bibinfo  {journal} {Review of
  Scientific Instruments}\ }\textbf {\bibinfo {volume} {86}},\ \href
  {https://doi.org/10.1063/1.4914834} {10.1063/1.4914834} (\bibinfo {year}
  {2015}),\ \bibinfo {note} {033109},\ \Eprint
  {https://arxiv.org/abs/https://pubs.aip.org/aip/rsi/article-pdf/doi/10.1063/1.4914834/15818465/033109\_1\_online.pdf}
  {https://pubs.aip.org/aip/rsi/article-pdf/doi/10.1063/1.4914834/15818465/033109\_1\_online.pdf}
  \BibitemShut {NoStop}%
\bibitem [{\citenamefont {Cram{\'e}r}(1999)}]{Cramer1999}%
  \BibitemOpen
  \bibfield  {author} {\bibinfo {author} {\bibfnamefont {H.}~\bibnamefont
  {Cram{\'e}r}},\ }\href@noop {} {\emph {\bibinfo {title} {Mathematical methods
  of statistics}}},\ Vol.~\bibinfo {volume} {9}\ (\bibinfo  {publisher}
  {Princeton university press},\ \bibinfo {year} {1999})\BibitemShut {NoStop}%
\bibitem [{\citenamefont {Rohatgi}\ and\ \citenamefont
  {Saleh}(2000)}]{Rohatgi2000}%
  \BibitemOpen
  \bibfield  {author} {\bibinfo {author} {\bibfnamefont {V.~K.}\ \bibnamefont
  {Rohatgi}}\ and\ \bibinfo {author} {\bibfnamefont {A.~M.~E.}\ \bibnamefont
  {Saleh}},\ }\href {https://doi.org/10.1002/9781118165676} {\emph {\bibinfo
  {title} {An introduction to probability and statistics}}}\ (\bibinfo
  {publisher} {John Wiley \& Sons},\ \bibinfo {year} {2000})\BibitemShut
  {NoStop}%
\bibitem [{\citenamefont {Helstrom}(1969)}]{Helstrom1969}%
  \BibitemOpen
  \bibfield  {author} {\bibinfo {author} {\bibfnamefont {C.~W.}\ \bibnamefont
  {Helstrom}},\ }\bibfield  {title} {\bibinfo {title} {Quantum detection and
  estimation theory},\ }\href {https://doi.org/10.1007/BF01007479} {\bibfield
  {journal} {\bibinfo  {journal} {Journal of Statistical Physics}\ }\textbf
  {\bibinfo {volume} {1}},\ \bibinfo {pages} {231} (\bibinfo {year}
  {1969})}\BibitemShut {NoStop}%
\bibitem [{\citenamefont {Holevo}(2011)}]{Holevo2011}%
  \BibitemOpen
  \bibfield  {author} {\bibinfo {author} {\bibfnamefont {A.}~\bibnamefont
  {Holevo}},\ }\href {https://books.google.co.uk/books?id=l7AIDhbWrTIC} {\emph
  {\bibinfo {title} {Probabilistic and Statistical Aspects of Quantum
  Theory}}},\ Publications of the Scuola Normale Superiore\ (\bibinfo
  {publisher} {Scuola Normale Superiore},\ \bibinfo {year} {2011})\BibitemShut
  {NoStop}%
\bibitem [{\citenamefont {Lee}\ \emph {et~al.}(2016)\citenamefont {Lee},
  \citenamefont {Kim}, \citenamefont {Cha},\ and\ \citenamefont
  {Moon}}]{Lee2016}%
  \BibitemOpen
  \bibfield  {author} {\bibinfo {author} {\bibfnamefont {S.~M.}\ \bibnamefont
  {Lee}}, \bibinfo {author} {\bibfnamefont {H.}~\bibnamefont {Kim}}, \bibinfo
  {author} {\bibfnamefont {M.}~\bibnamefont {Cha}},\ and\ \bibinfo {author}
  {\bibfnamefont {H.~S.}\ \bibnamefont {Moon}},\ }\bibfield  {title} {\bibinfo
  {title} {Polarization-entangled photon-pair source obtained via type-ii
  non-collinear spdc process with ppktp crystal},\ }\href
  {https://doi.org/10.1364/OE.24.002941} {\bibfield  {journal} {\bibinfo
  {journal} {Opt. Express}\ }\textbf {\bibinfo {volume} {24}},\ \bibinfo
  {pages} {2941} (\bibinfo {year} {2016})}\BibitemShut {NoStop}%
\bibitem [{Coa()}]{Coarse}%
  \BibitemOpen
  \href@noop {} {}\bibinfo {note} {This is a coarse and simplified estimate as
  it does not take into consideration more refined technical details, such as
  the birifrangence or the periodic poling of the crystal for type-II SPDC, the
  phase-matching relation between the aperture angle, the tilting of the
  optical axes and the refractive indices, but nevertheless it gives an idea of
  the order of magnitude of $\Delta k$.}\BibitemShut {Stop}%
\bibitem [{\citenamefont {Devaux}\ \emph {et~al.}(2020)\citenamefont {Devaux},
  \citenamefont {Mosset}, \citenamefont {Moreau},\ and\ \citenamefont
  {Lantz}}]{Devaux2020}%
  \BibitemOpen
  \bibfield  {author} {\bibinfo {author} {\bibfnamefont {F.}~\bibnamefont
  {Devaux}}, \bibinfo {author} {\bibfnamefont {A.}~\bibnamefont {Mosset}},
  \bibinfo {author} {\bibfnamefont {P.-A.}\ \bibnamefont {Moreau}},\ and\
  \bibinfo {author} {\bibfnamefont {E.}~\bibnamefont {Lantz}},\ }\bibfield
  {title} {\bibinfo {title} {Imaging spatiotemporal hong-ou-mandel interference
  of biphoton states of extremely high schmidt number},\ }\href
  {https://doi.org/10.1103/PhysRevX.10.031031} {\bibfield  {journal} {\bibinfo
  {journal} {Phys. Rev. X}\ }\textbf {\bibinfo {volume} {10}},\ \bibinfo
  {pages} {031031} (\bibinfo {year} {2020})}\BibitemShut {NoStop}%
\bibitem [{\citenamefont {López-Durán}\ and\ \citenamefont
  {Rosas-Ortiz}(2022)}]{LopezDuran2022}%
  \BibitemOpen
  \bibfield  {author} {\bibinfo {author} {\bibfnamefont {J.}~\bibnamefont
  {López-Durán}}\ and\ \bibinfo {author} {\bibfnamefont {O.}~\bibnamefont
  {Rosas-Ortiz}},\ }\bibfield  {title} {\bibinfo {title} {Exact solutions for
  vector phase-matching conditions in nonlinear uniaxial crystals},\ }\bibfield
   {journal} {\bibinfo  {journal} {Symmetry}\ }\textbf {\bibinfo {volume}
  {14}},\ \href {https://doi.org/10.3390/sym14112272} {10.3390/sym14112272}
  (\bibinfo {year} {2022})\BibitemShut {NoStop}%
\bibitem [{\citenamefont {Giovannetti}\ \emph {et~al.}(2006)\citenamefont
  {Giovannetti}, \citenamefont {Lloyd},\ and\ \citenamefont
  {Maccone}}]{Giovannetti2006}%
  \BibitemOpen
  \bibfield  {author} {\bibinfo {author} {\bibfnamefont {V.}~\bibnamefont
  {Giovannetti}}, \bibinfo {author} {\bibfnamefont {S.}~\bibnamefont {Lloyd}},\
  and\ \bibinfo {author} {\bibfnamefont {L.}~\bibnamefont {Maccone}},\
  }\bibfield  {title} {\bibinfo {title} {Quantum metrology},\ }\href
  {https://doi.org/10.1103/PhysRevLett.96.010401} {\bibfield  {journal}
  {\bibinfo  {journal} {Phys. Rev. Lett.}\ }\textbf {\bibinfo {volume} {96}},\
  \bibinfo {pages} {010401} (\bibinfo {year} {2006})}\BibitemShut {NoStop}%
\end{thebibliography}%

\end{document}